\documentclass[aps,prx,twocolumn,superscriptaddress,showpacs,longbibliography,floatfix]{revtex4-2}
\usepackage{xcolor}
\usepackage{graphicx}
\usepackage{amsmath,amssymb,bm,ulem}
\usepackage{physics}
\usepackage{hyperref}
\usepackage{placeins}

\begin{document}

\title{Measurement-Only Dynamical Phase Transitions in Spin-1 Chains}

\author{Kemal Aziz}
\affiliation{Department of Physics and Astronomy, Center for Materials Theory, Rutgers University, Piscataway, NJ 08854, USA}
\author{Haining Pan}
\affiliation{Department of Physics and Astronomy, Center for Materials Theory, Rutgers University, Piscataway, NJ 08854, USA}
\author{J. H. Pixley}
\affiliation{Department of Physics and Astronomy, Center for Materials Theory, Rutgers University, Piscataway, NJ 08854, USA}
\affiliation{Center for Computational Quantum Physics, Flatiron Institute, New York, New York 10010, USA}

\begin{abstract}
We study measurement-only dynamics in an interacting spin-1 chain, using forced, normalized local projections rather than Born-rule-sampled measurement outcomes. Starting from the nearest-neighbor (NN) Affleck--Kennedy--Lieb--Tasaki (AKLT) Haldane symmetry-protected topological state, we show that competing local projectors drive this order toward three distinct dynamical phases: a featureless topologically trivial product state, an explicitly dimerized valence-bond-solid state, and a topologically trivial next-nearest-neighbor (NNN) AKLT state consisting of two interwoven Haldane chains. Using matrix product state and exact statevector simulations, we identify continuous transitions among these phases and characterize them using nonlocal string order parameters, local dimer order, and entanglement scaling.
For the NN--NNN AKLT model, finite-size signatures are consistent with an intervening extended critical regime rather than a single direct transition. These results extend studies of monitored topology to a bosonic SPT and show that local projections can control both topological and valence-bond-solid order.
\end{abstract}

\maketitle

\section{Introduction}
\label{sec:introduction}
The Affleck-Kennedy-Lieb-Tasaki (AKLT) chain is a canonical one-dimensional model of a symmetry-protected topological (SPT) phase with an exactly known ground-state wavefunction~\cite{affleckRigorousResultsValencebond1987,affleckValenceBondGround1988}.
It is a spin-1 chain with bilinear and biquadratic nearest-neighbor (NN) couplings in a fixed ratio, by tuning this ratio the model can pass between several phases that include the topological AKLT state,   a valence-bond solid (VBS) with dimer coverings that break translational symmetry,  a ferromagnetic state, and  a nematic state. 
The AKLT ground state is gapped with exponentially decaying spin correlations, as Haldane conjectured for integer-spin Heisenberg chains~\cite{haldaneNonlinearFieldTheory1983}. 
The resulting Haldane phase is protected by the $\mathbb{Z}_2\times\mathbb{Z}_2$ symmetry generated by $\pi$ spin rotations about two orthogonal axes~\cite{pollmannEntanglementSpectrumTopological2010}. 
This order is stable to symmetry-preserving perturbations and has been identified on noisy quantum hardware~\cite{azsesIdentificationSymmetryProtectedTopological2020}. The two-dimensional AKLT state on the honeycomb lattice is a universal resource for measurement-based quantum computation~\cite{weiAffleckKennedyLiebTasakiStateHoneycomb2011,miyakeQuantumComputationalCapability2011}. For spin-1 chains in the Haldane phase, a smaller magnitude of the string-order parameter increases the measurement overhead required to implement a target single-qubit gate~\cite{yangMeasurementbasedQuantumComputation2026}.
In the classification of gapped phases it is a \textit{bosonic} SPT phase, in the sense that its degrees of freedom are spins, whose operators on different sites commute, rather than fermionic modes~\cite{chenClassificationGappedSymmetric2011b}.

For a fixed protecting on-site symmetry, gapped one-dimensional SPT phases are distinguished by a discrete topological index. This index is invariant under symmetry-preserving local unitary circuits whose gate range and depth are bounded as the system size increases. Such a circuit therefore cannot connect states belonging to distinct SPT phases. The normalized projections studied here are nonunitary and are not subject to this unitary circuit constraint. The same bounded-depth restriction underlies the Lieb-Robinson result that short depth local unitary circuits cannot generate long-range entanglement (LRE) from a product state~\cite{chenLocalUnitaryTransformation2010c,liebFiniteGroupVelocity1972,piroliQuantumCircuitsAssisted2021}.
Circuits that break the protecting symmetry are not subject to this obstruction, because symmetry breaking allows distinct gapped SPT states to be connected.

By moving away from the requirement of unitary evolution, we can  get around some of these restrictions. For example, measurement-based imaginary time evolution (MITE), e.g. $|\psi(\tau)\rangle = e^{-H\tau}|\psi\rangle/||e^{-H\tau}|\psi\rangle||$ for real $\tau$
can connect distinct SPT phases, or an SPT phase and a topologically trivial phase, in finite depth, unlike symmetry-preserving finite-depth local unitary evolution~\cite{maoMeasurementbasedDeterministicImaginary2023,nishiImplementationQuantumImaginarytime2021,chenEfficientPreparationAKLT2024a,piroliQuantumCircuitsAssisted2021}. This is because nonunitary operations are not necessarily symmetry-preserving or invertible~\cite{nielsenQuantumComputationQuantum2010}. This means they lie outside the reversible, symmetry-preserving equivalence between SPT phases~\cite{piroliQuantumCircuitsAssisted2021}. This partly motivates our study of transitions between distinct phases driven by competing nonunitary projectors.

Dynamics generated by forced measurements are closely related to outcome-conditioned monitored dynamics that has been recently studied in free-fermion topological systems. In that setting, monitored circuits have been used to classify topological modes under measurement~\cite{panTopologicalModesMonitored2025,bhuiyanFreefermionDynamicsMeasurements2026,oshimaTopologySpectrumMeasurementInduced2025,xiaoSymmetryTopologyMonitored2026}. More generally, measurement protocols with outcome-conditioned classical feedback have enabled the demonstration of topological order and the preparation of target states. 
In a similar fashion, fusion measurements combined with shallow-depth unitaries can deterministically prepare the NN AKLT state~\cite{smithDeterministicConstantDepthPreparation2023e}. 

The Haldane phase, for which the AKLT model provides an exactly known ground-state wavefunction, has also been realized experimentally, first in quasi-one-dimensional spin-1 antiferromagnets~\cite{buyersExperimentalEvidenceHaldane1986,regnaultInelasticneutronscatteringStudySpin1994a}, and more recently via quantum state preparation on noisy intermediate-scale quantum (NISQ) hardware~\cite{chenHighfidelityRealizationAKLT2023,edmundsConstructingSpin1Haldane2025}. More broadly, measurement-based protocols have been used to prepare valence-bond-solid at shallow circuit depth~\cite{murtaPreparingValencebondsolidStates2023}, Greenberger-Horne-Zeilinger (GHZ) cat states~\cite{baumerEfficientLongRangeEntanglement2024}, and toric-code ground states~\cite{luMeasurementShortcutLongRange2022,aguadoCreationManipulationDetection2008}. Measurement-induced transitions involving topological or symmetry-protected order have also been studied in a variety of settings~\cite{lavasaniMeasurementinducedTopologicalEntanglement2021,sangMeasurementprotectedQuantumPhases2021,klockeTopologicalOrderEntanglement2022,morral-yepesDetectingStabilizingMeasurementinduced2023,kunoEmergenceSymmetryProtected2022,yuGaplessSymmetryprotectedTopological2026,zhuSteadystatePhasesLongrange2026,tantivasadakarnLongRangeEntanglementMeasuring2024}.

In the following manuscript we study the dynamical competition between several ground states of the spin 1 chain with nearest and next nearest neighbor bilinear and biquadratic interactions. The relevant competing states we consider are the AKLT state against, (1) a topologically trivial product state under strong single-ion anisotropy~\cite{chenClassificationGappedSymmetric2011b}, (2) a spontaneously dimerized phase at a different bilinear-biquadratic ratio~\cite{fathSearchNondimerizedQuantum1995,schollwockOnsetIncommensurabilityValencebondsolid1996,patiComparativeStudyPhase1997}, and (3) an NNN AKLT phase of two interleaved AKLT chains under next-nearest-neighbor (NNN) exchange~\cite{kolezhukFirstOrderTransition1996,kolezhukVariationalDensitymatrixRenormalizationgroup1997,kolezhukConnectivityTransitionFrustrated2002,pixleyFrustrationMulticriticalityAntiferromagnetic2014}.
We let the NN AKLT projector compete with three local noncommuting projectors, one favoring each of these states. The first projects every site independently onto $S_i^z=0$ and favors a featureless large-anisotropy product state with no residual multisite structure. 
The second favors an explicitly dimerized valence-bond-solid state, with one-site translation symmetry broken by the choice of projected bonds. 
The third alternative favors the NNN AKLT ground state, which decouples into two independent, interwoven AKLT chains living on the even and odd sublattices. We refer to it as the topologically trivial NNN AKLT regime.
We interpolate from the NN AKLT projector to the NNN AKLT projector, respecting the $\mathbb{Z}_2\times\mathbb{Z}_2$ symmetry embedded in the full SU(2) spin rotational symmetry, and define the NN AKLT state as the topologically nontrivial phase and NNN AKLT state as the topologically trivial phase following the Z2 topological index.~\cite{pollmannEntanglementSpectrumTopological2010,chenClassificationGappedSymmetric2011b}
It nevertheless retains the structure of two interleaved AKLT chains and nonzero short-range entanglement, unlike the unentangled product state.

With these competing states defined, measurement-only dynamics does not necessarily generate rich entanglement structure: if the competing projectors act on single sites or otherwise commute with one another, repeated projection simply collapses the system onto an area law entangled product state~\cite{ippolitiEntanglementPhaseTransitions2021}. Growing string order and entanglement structure instead requires competing, noncommuting projectors, which is the regime we study here.
We find three distinct transitions out of the NN AKLT Haldane phase under such competition: it can be destroyed in favor of the featureless product state without generating any local symmetry-breaking order; it can be converted into explicit dimer VBS order; or, when competed directly against the NNN AKLT projector, it can give way to an intervening critical regime separating the two AKLT phases. We diagnose all three models using nonlocal string order parameters, local dimer order, and entanglement scaling.

The remainder of the paper is organized as follows: in Sec.~\ref{sec:models}, we introduce the three measurement-only spin-1 protocols and the forced projectors that define them. In Sec.~\ref{sec:results}, we use string order, dimer order, entanglement, and finite-size scaling to characterize the resulting measurement-induced transitions. We discuss broader implications in Sec.~\ref{sec:discussion} and summarize our results and outlook in Sec.~\ref{sec:conclusion}. Appendix~\ref{sec:dynamical_exponent} presents the dynamical-exponent collapses. Appendix~\ref{sec:random_product_fidelity} examines convergence to the deterministic target states. Appendix~\ref{sec:analytic_calculations} analytically computes the string order reference values and periodic-chain entanglement limits for the NN and NNN AKLT states. Appendix~\ref{sec:finite_size_scaling} summarizes the order-parameter finite-size scaling, and Appendix~\ref{sec:bond_dimension} discusses bond dimension diagnostics and finite-size cost scaling. Appendix~\ref{sec:nn_nnn_saturation} analyzes relaxation and saturation in the NN--NNN AKLT circuit, and Appendix~\ref{sec:nn_nnn_renyi_entanglement} examines the R\'enyi-index dependence of entanglement in the NN--NNN critical phase.

\begin{figure*}[htbp]
    \includegraphics[width=7in]{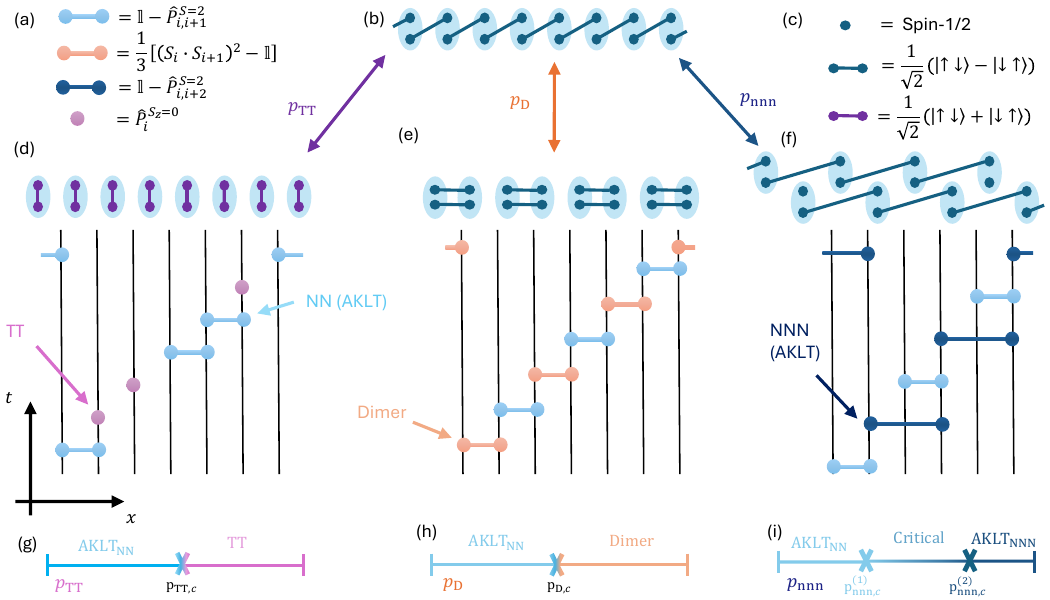}
    \caption[]{
    Overview of the measurement-only spin-1 protocols.
    (a) Local projectors used in the three protocols: the NN AKLT projector (light blue), dimer VBS projector (orange), NNN AKLT projector (dark blue), and single-site $S_i^z=0$ projector (purple). (b) Virtual-spin construction of the NN AKLT state. The solid points denote virtual spin-1/2 degrees of freedom paired into singlets; the ovals project each pair of virtual spins onto the physical spin-1 subspace. (c) Conventions for the virtual spin-1/2 degrees of freedom and their singlet and $S^z=0$ triplet pairings. (d) Circuit diagram of the SPT-to-TT model. For each qutrit $i$, the single-site $S_{i}^{z}=0$ projector is applied on $i$ with probability $p_{\mathrm{TT}}$, or the NN AKLT projector is applied on $(i,i+1)$ with probability $1-p_{\mathrm{TT}}$. The TT product state is depicted above the circuit and is the final state for $p_{\mathrm{TT}}=1$. (e) Circuit diagram of the SPT-to-dimer model. On odd bonds $(i,i+1)$, the dimer VBS projector is applied with probability $p_{\mathrm{D}}$; with probability $1-p_{\mathrm{D}}$, the NN AKLT projector is applied instead. The dimer VBS state is depicted above the circuit and is the final state for $p_{\mathrm{D}}=1$. (f) Circuit diagram of the NN--NNN AKLT model. For each qutrit $i$, the NNN AKLT projector is applied on $(i,i+2)$ with probability $p_{\mathrm{nnn}}$, or the NN AKLT projector is applied on $(i,i+1)$ with probability $1-p_{\mathrm{nnn}}$. The NNN AKLT state is depicted above the circuit and is the final state for $p_{\mathrm{nnn}}=1$. (g) Phase diagram with $p_{\mathrm{TT},c}$ marking the separation between the NN AKLT and TT phases. (h) Phase diagram with $p_{\mathrm{D},c}$ marking the separation between the NN AKLT and dimer phases. (i) Phase diagram with $p_{\mathrm{nnn},c}^{(2)}$ and $p_{\mathrm{nnn},c}^{(1)}$ marking, respectively, the NN AKLT--critical and critical--NNN AKLT boundaries.
    }
    \label{fig:AKLT_circuit_diagram}
\end{figure*}

\section{Models} 
\label{sec:models}
We construct three models to study how an interacting bosonic Haldane SPT phase responds to competing projector dynamics.
In each model, the local NN AKLT projector competes with an incompatible local projector favoring a featureless product state, an explicitly dimerized VBS state, or the interwoven NNN AKLT structure.
Selectively enforcing projectors onto different symmetry sectors gives access to the preparation of several area law wavefunctions: the Haldane SPT, NNN AKLT, topologically trivial product, and explicitly dimerized VBS states.
We refer to the competing-local-projector dynamics studied here as a forced normalized local-projection protocol.
At each local update, the relevant probability parameter first selects which competing projector (i.e. ``filter'') is applied. Once selected, that projection branch is forced rather than sampled as a measurement outcome following the Born rule; the selected filter is applied and the resulting state is then renormalized~\cite{nielsenQuantumComputationQuantum2010}.  
Randomness therefore enters through choices of projective measurement, whereas Born-rule sampling of measurement outcomes is not part of the protocol. Here ``measurement-only'' names this projector-only dynamical framework and does not assert a physical or hardware realization.

We consider MITE-inspired measurement-only dynamics on a chain of $L$ spin-1 degrees of freedom (qutrits) arranged in the staircase circuit geometries shown in Figs.~\ref{fig:AKLT_circuit_diagram}(d)--(f). The dynamics is defined in terms of local forced projectors: at each update step, one of several competing local filters is applied and the state is renormalized. In simulations, these forced measurements are implemented using normalized ITE,
\begin{equation}
|\psi\rangle \mapsto \lim_{\tau \rightarrow \infty} \frac{e^{-\hat{\mathcal{H}}\tau}|\psi\rangle}{\lVert e^{-\hat{\mathcal{H}}\tau}|\psi\rangle\rVert}.
\label{eq:forced_measurement_eq}
\end{equation}
This projects onto the ground-state subspace of the operator $\hat{\mathcal{H}}$.
These spin-1 protocols are motivated by MITE, which uses measurement and feedback to approximate nonunitary imaginary-time filters by the choice of $\hat{\mathcal{H}}$, including filters for AKLT-state preparation~\cite{maoMeasurementbasedDeterministicImaginary2023,chenEfficientPreparationAKLT2024a}. Starting from $|\psi_0\rangle$, we define one timestep as a single local projector application, so a left-to-right sweep over $i=1,\ldots,L$ contains $L$ timesteps. In every model, the light-blue NN AKLT projector in Fig.~\ref{fig:AKLT_circuit_diagram}(a) stabilizes the reference SPT phase and competes with one of the following projectors:
\begin{itemize}
    \item A single-site $S_{i}^{z}=0$ projector favoring a topologically trivial (TT) product state, applied with probability $p_{\mathrm{TT}}$, depicted as the purple dot in Fig.~\ref{fig:AKLT_circuit_diagram}(a) and (d).
    \item A dimer VBS projector on odd bonds, applied with probability $p_{\mathrm{D}}$, shown as an orange bond in Fig.~\ref{fig:AKLT_circuit_diagram}(a) and (e).
    \item An NNN AKLT projector applied with probability $p_{\mathrm{nnn}}$, depicted as a dark blue bond that stretches across three sites in Fig.~\ref{fig:AKLT_circuit_diagram}(a) and (f).
\end{itemize} 
Throughout the paper, an endpoint of a phase diagram means one of the pure-projector limits obtained by setting the tuning probability to either zero or one. Among the protocols studied here, the AKLT and dimer transitions are generated entirely by SU(2)-invariant total-spin projectors. The topologically trivial protocol is different: its competing single-ion $S_i^z=0$ filter explicitly breaks SU(2) spin-rotation symmetry down to the subgroup preserving the chosen quantization axis.
In each model, the NN AKLT projector is applied with complementary probability ($1-p_{\alpha}$). For the NN--NNN protocol, we denote this probability by $p_{\mathrm{nn}}=1-p_{\mathrm{nnn}}$. All numerical simulations in this work use periodic boundary conditions (PBC), with site indices understood modulo $L$.
Thus the NN AKLT endpoint occurs when the NN AKLT projector is selected with probability one, while the competing endpoints are the TT product state at $p_{\mathrm{TT}}=1$, the explicitly dimerized VBS at $p_{\mathrm{D}}=1$, and the NNN AKLT state at $p_{\mathrm{nnn}}=1$.

For each model, we denote the corresponding projector acting on the relevant set of sites $\{i\}$ by
\begin{equation}
\hat{P}^{\mathrm{model}}_{\{i\}}=\lim_{\tau \rightarrow \infty} e^{-\hat{\mathcal{H}}_{\{i\}}^{\mathrm{model}}\tau},
\end{equation}
and depicted in Fig.~\ref{fig:AKLT_circuit_diagram}(a)
A common parent Hamiltonian containing the NN, NNN, and single-ion terms relevant to the projector limits studied below is
\begin{equation}
{
\begin{aligned}
\hat{\mathcal{H}}
=\sum_{i=1}^{L}\Big[
&J_{1}\,\mathbf{S}_{i}\cdot\mathbf{S}_{i+1}
+K_{1}\,(\mathbf{S}_{i}\cdot\mathbf{S}_{i+1})^{2}+D(S_{i}^{z})^{2}
\\
&+J_{2}\,\mathbf{S}_{i}\cdot\mathbf{S}_{i+2}
+K_{2}\,(\mathbf{S}_{i}\cdot\mathbf{S}_{i+2})^{2}
\Big].
\end{aligned}}
\label{eq:parent_hamiltonian}
\end{equation}
Here $\mathbf{S}_{i}$ is the spin-1 operator on site $i$, with site labels understood modulo $L$ for PBC. The couplings $J_{1}$ and $K_{1}$ set the NN bilinear and biquadratic interactions, while $J_{2}$ and $K_{2}$ set the corresponding NNN interactions. The coefficient $D$ is the single-ion anisotropy whose large positive limit favors the TT product state. The NN and NNN AKLT limits are obtained by choosing the bilinear-biquadratic ratio $K_{r}=J_{r}/3$ on the active bond family $r=1$ or $r=2$, respectively, with all competing terms set to zero up to an additive constant and an overall positive scale.
Depending on its couplings, this parent Hamiltonian supports ferromagnetic, Haldane, dimerized, and other phases. We use its local terms below to define the competing projectors.
\subsection{Nearest-Neighbor AKLT Ground State}
The NN AKLT limit of Eq.~\eqref{eq:parent_hamiltonian} is obtained by setting $J_1>0$, $K_1=J_1/3$, $J_2=K_2=0$, and $D=0$, with $J_1$ fixing the overall energy scale. Up to an additive constant and an overall positive scale, the resulting Hamiltonian is

\begin{equation}
\hat{\mathcal{H}}_{\mathrm{AKLT}} = \sum_{j=1}^{L} \hat{P}_{j,j+1}^{S=2}
\label{eq:AKLT_ground_state}
\end{equation}
for the periodic chain, where 
\begin{equation}
\hat{P}_{j,j+1}^{S=2} = \frac{1}{2}[\mathbf{S}_{j} \cdot \mathbf{S}_{j+1} + \frac{1}{3}(\mathbf{S}_{j} \cdot \mathbf{S}_{j+1})^{2} + \frac{2}{3}\mathbb{I}]
\label{eq:NN_AKLT_projector}
\end{equation}
is the projector onto the $S=2$ subspace for qutrits $j$ and $j+1$. We define the forced measurement projector  
\begin{equation}
\hat{P}_{j,j+1}^{\mathrm{nn}}
\;\equiv\;
\lim_{\tau\to\infty} e^{-\tau\, \hat P_{j,j+1}^{S=2}}
= \mathbb{I} - \hat P_{j,j+1}^{S=2}.
\label{eq:NN_AKLT_projection}
\end{equation}
as represented by the light-blue NN AKLT bond symbol in Fig.~\ref{fig:AKLT_circuit_diagram}(a) and by the light-blue updates in Figs.~\ref{fig:AKLT_circuit_diagram}(d)--(f). In the limit of NN-only measurements, we have 
\begin{equation}
\lim_{n \rightarrow \infty} \frac{(\prod_{j=1}^{L} \hat{P}_{j,j+1}^{\mathrm{nn}})^{n}|\psi_0\rangle}{\lVert(\prod_{j=1}^{L} \hat{P}_{j,j+1}^{\mathrm{nn}})^{n}|\psi_0\rangle\rVert} = |\mathrm{AKLT}\rangle,
\end{equation}
where $\hat{\mathcal{H}}_{\mathrm{AKLT}}|\mathrm{AKLT}\rangle = 0$ because $\hat{P}^{S=2}_{j,j+1}|\mathrm{AKLT}\rangle = 0$. The projectors on adjacent pairs of sites do not commute, $[\hat{P}_{j-1,j}^{S=2},\hat{P}_{j,j+1}^{S=2}] \neq 0$, even though the ground state is a common eigenstate of all the projection operators. We sequentially apply the normalized single-bond filters $e^{-\tau \hat P^{S=2}_{j,j+1}}$ and take $\tau\to\infty$ on each bond. Although the projectors do not commute, the AKLT state is a simultaneous zero-energy state, so repeated alternating application converges to $|\mathrm{AKLT}\rangle$. In the PBC simulations used in the main text, the product extends over all bonds $j=1,\ldots,L$, with $j+1$ understood modulo $L$, and produces the nondegenerate AKLT ground state. For OBC, the product stops at $j=L-1$, leaving two uncoupled spin-1/2 edge modes and a fourfold ground-state degeneracy; the final state can therefore retain memory of the edge sector selected by the initial state or boundary protocol.

We show that this preparation scheme accurately produces the desired product state wavefunction, when it is known exactly in 
Appendix~\ref{sec:random_product_fidelity}.

\subsection{Next-Nearest-Neighbor AKLT Ground State}
\label{sec:nnn_aklt_gs}
The NNN AKLT limit of Eq.~\eqref{eq:parent_hamiltonian} is obtained by setting $J_{2}>0$, $K_{2}=J_{2}/3$, $J_{1}=K_{1}=0$, and $D=0$, with $J_{2}$ fixing the overall energy scale. Up to an additive constant and an overall positive scale, this gives the PBC NNN AKLT Hamiltonian
\begin{equation}
{\hat{\mathcal{H}}_{\mathrm{nnn}}^{\mathrm{AKLT}} = \sum_{i=1}^{L} \hat{P}_{i,i+2}^{S=2}.}
\label{eq:nnn_aklt_hamiltonian}
\end{equation}

We separate the next-nearest-neighbor projectors into even and odd sums, so that each acts on disjoint sets of qutrits. Physically, this defines two interwoven AKLT chains on the even and odd sites.
We define the forced measurement projector   
\begin{equation}
\hat{P}_{j,j+2}^{\mathrm{nnn}}
\;\equiv\;
\lim_{\tau\to\infty} e^{-\tau\, \hat P_{j,j+2}^{S=2}}
= \mathbb{I} - \hat P_{j,j+2}^{S=2}.
\label{eq:next-nearest-neighbor-aklt-projector}
\end{equation}
as shown by the dark-blue NNN AKLT bond symbol in Fig.~\ref{fig:AKLT_circuit_diagram}(a) and by the NNN circuit geometry in Fig.~\ref{fig:AKLT_circuit_diagram}(f). In the limit of NNN-only measurements on the periodic chain, with \(j+2\) understood modulo \(L\), the dynamics projects onto two decoupled AKLT chains supported on the even and odd sublattices:
\begin{equation}
\begin{aligned}
&\lim_{n \rightarrow \infty}
\frac{
  \big(\prod_{j=1}^{L} \hat{P}_{j,j+2}^{\mathrm{nnn}}\big)^{n}
  |\psi_0\rangle
}{
  \big\|\big(\prod_{j=1}^{L} \hat{P}_{j,j+2}^{\mathrm{nnn}}\big)^{n}
  |\psi_0\rangle\big\|
}
\\
&\qquad =
|\mathrm{AKLT}\rangle_{\mathrm{even}}
\otimes
|\mathrm{AKLT}\rangle_{\mathrm{odd}}.
\end{aligned}
\end{equation}
The factorized endpoint above is two interwoven AKLT chains. With only the diagonal on-site SO(3) symmetry, or its $D_2$ subgroup, acting on the physical spin-1 sites, the one-dimensional Haldane index is $\mathbb{Z}_2$, so the two copies stack to the trivial diagonal-symmetry SPT class: the two edge spin-$1/2$ projective representations can combine into a linear representation~\cite{pollmannEntanglementSpectrumTopological2010,chenClassificationGappedSymmetric2011b}. It remains physically distinct in this protocol, diagnosed here by the NNN string order and by its doubled entanglement structure; these observables distinguish the endpoint regimes but are not, by themselves, complete invariants of the diagonal-symmetry SPT class. The symmetry distinction is as follows. At the isolated NNN-only point, one may assign independent protecting symmetries to the two decoupled sublattices, $D_2^{\mathrm{even}}\times D_2^{\mathrm{odd}}$; with this enlarged symmetry, each sublattice is an AKLT chain with its own projective edge representation. This enlarged symmetry is not preserved by the mixed NN--NNN dynamics. A nearest-neighbor AKLT projector acts on one even and one odd site, and is invariant only under the diagonal action in which both sites are rotated by the same $D_2$ element, not under independent even- and odd-sublattice rotations. Thus for any nonzero NN-projector rate, the symmetry of the protocol is reduced to the diagonal subgroup $D_2^{\mathrm{diag}}=\{(g,g)\}$. Under this diagonal symmetry, the two Haldane chains are stacked: the even- and odd-chain edge spin-$1/2$ modes at the same boundary can be paired by a symmetry-preserving local coupling, so the NNN AKLT endpoint lies in the trivial diagonal-symmetry SPT class. The corresponding deterministic endpoint preparation is checked by the fidelity diagnostics in Appendix~\ref{sec:random_product_fidelity}.

\subsection{Dimer VBS Ground State}
The dimer projector projects onto the dimer VBS ground state of the Hamiltonian
\begin{equation}
\hat{\mathcal H}_{\mathrm{dimer}}
=
-\sum_{i=1}^{L/2}
\big(\mathbf S_{2i-1}\cdot\mathbf S_{2i}\big)^2.
\label{eq:Dimer_gs_Hamiltonian}
\end{equation}
On the active odd bonds $(2i-1,2i)$ we set $J_{1}=0$ and $K_{1}=-1$, while the inactive even bonds have $J_{1}=K_{1}=0$. All NNN and single-ion terms are set to zero, $J_{2}=K_{2}=D=0$. The alternating choice of active bonds explicitly selects one of the two translation-related dimer patterns.

The dimer projector maps the bond between sites $j$ and $j+1$ onto a singlet, or dimer, and is defined by
\begin{equation}
{\color{black}
\hat{P}_{j,j+1}^{\mathrm{D}} = \lim_{\tau \rightarrow \infty} e^{\tau[(\mathbf{S}_{j} \cdot \mathbf{S}_{j+1})^{2}-4]}=\frac{1}{3}[(\mathbf{S}_{j} \cdot\mathbf{S}_{j+1})^{2} - \mathbb{I}].}
\label{eq:single_bond_dimer_projector}
\end{equation}
For two spin-1 particles in the $S^{z}$ basis, $\hat{P}_{j,j+1}^{\mathrm{D}}$ projects onto the unique singlet state.

\begin{equation}
{\color{black}
|D\rangle_{j,j+1} = \frac{1}{\sqrt{3}}(|1,-1\rangle+|-1,1\rangle-|0,0\rangle).}
\label{eq:dimer_state}
\end{equation}
For an even-length chain with the chosen odd-bond dimerization, the dimer VBS ground state of Eq.~\eqref{eq:Dimer_gs_Hamiltonian} is
\begin{equation}
{\color{black}
|\psi_{\mathrm{dimer}}^{\mathrm{PBC}}\rangle = \bigotimes_{j=1}^{L/2} |D\rangle_{2j-1,2j}.}
\label{eq:dimer_VBS_gs}
\end{equation}
Here and below, ``symmetry-breaking'' refers to the one-site translation symmetry broken explicitly by the odd-bond Hamiltonian and projector pattern, not to spontaneous selection between symmetry-related dimer sectors.
In the limit of dimer-only measurements on the odd bonds, we have
\begin{equation}
\frac{
\prod_{j=1}^{L/2} \hat{P}_{2j-1,2j}^{\mathrm{D}}\ket{\psi_{0}}
}{
\left\|
\prod_{j=1}^{L/2}\hat{P}_{2j-1,2j}^{\mathrm{D}}\ket{\psi_{0}}
\right\|
}
=\ket{\psi_{\mathrm{dimer}}^{\mathrm{PBC}}}.
\label{eq:dimer_projection}
\end{equation}
 
\subsection{Topologically Trivial Product State}
The final competing projector favors the topologically trivial (TT) product-state ground state of $\hat{\mathcal{H}}=\hat{\mathcal{H}}_{\mathrm{ion}} + \hat{\mathcal{H}}_{\mathrm{AKLT}}$. The on-site $\mathbb{Z}_2\times\mathbb{Z}_2$ symmetry (e.g., $\pi$ rotations about orthogonal spin axes) that protects and distinguishes the neighboring AKLT SPT phase is preserved by our measurement dynamics.
\begin{equation}
\hat{\mathcal{H}}_{\mathrm{ion}} = D \sum_{i}(S_{i}^{z})^{2},
\label{eq:anisotropy}
\end{equation}
with $D>0$. As a result, 
$\hat{\mathcal{H}}_{\mathrm{ion}}$'s ground state has 
$S^{z} = 0$  on each qutrit $i$. We define the forced measurement projector
\begin{equation}
{\hat{P}_{i}^{S^{z}=0}} = \lim_{\tau \rightarrow \infty} e^{-D(S_i^z)^{2}\tau}
\label{eq:trivial_projection}
\end{equation}
where $\hat{P}_{i}^{S^{z}=0}$ projects onto the $S^z=0$ subspace on qutrit $i$. When only this projector is applied, the wavefunction is projected onto the ground state $|\psi\rangle=|0\rangle^{\otimes L}$.
We demonstrate this procedure produces the exact 
fidelity 
in Appendix~\ref{sec:random_product_fidelity}.

\section{Results}
\label{sec:results}
We present results for the three measurement-only spin-1 models summarized in Figs.~\ref{fig:AKLT_circuit_diagram}(d)--(i) that all study transitions out of the AKLT wavefunction. The competing projector favors, respectively, the symmetry-preserving product state $|0\rangle^{\otimes L}$, an explicitly selected odd-bond dimer VBS, or the topologically trivial NNN AKLT state. We tune the relative projector probabilities to locate the resulting phase transitions. The SPT--TT and SPT--dimer models are amenable to simulations with MPS, whereas the NN--NNN AKLT model has far too much entanglement and we find no benefit to using MPS and is simulated with exact statevector evolution, instead. This we attribute to the existence of a critical phase in the phase diagram. After each MPS update, we set the truncation error to $\epsilon=10^{-10}$ and allow the bond dimension $\chi$ to grow as needed; Appendix~\ref{sec:bond_dimension} analyzes this growth. 

The simulations are initialized with a random MPS,
\begin{equation}
|\psi_0\rangle =
\sum_{s_1,\ldots,s_L}
A_1^{(s_1)}A_2^{(s_2)}\cdots A_L^{(s_L)}
|s_1s_2\cdots s_L\rangle,
\end{equation}
where $s_i\in\{-1,0,+1\}$ labels the local spin-1 basis and the tensors $A_i^{(s_i)}$ are chosen randomly before normalizing the state. The purpose of this choice is simply to give the forced-projector dynamics nonzero overlap with the relevant target subspaces. For PBCs, the steady state in the pure-projector limits is expected to be independent of this initial random MPS, because the periodic AKLT ground state is nondegenerate and the forced dynamics converges to the unique ground state whenever the initial state has nonzero overlap with it. For OBCs, by contrast, the AKLT ground space is four-fold degenerate due to the two spin-$1/2$ edge modes, so the final edge sector can retain memory of the initial state.
 Each realization begins from an independently chosen random initial MPS and evolves under an independently sampled sequence of projector choices. When spatial self-averaging is used, string order observables are first averaged over equivalent placements of the string operator within a realization and then averaged over realizations; halfcut entanglement in PBC is averaged over sampled cut locations before the trajectory average. We use $\mathcal{O}(10^{3})$ trajectories for each figure and use an overline (e.g. $\overline{A}$) to denote an average over independent circuit realizations.

For the steady-state order-parameter collapses used throughout this section, we use the common finite-size scaling form
\begin{equation}
\overline{A}(L,p)
=L^{-\Delta_A}\,
\mathcal{F}_{A}\!\left[(p-p_c)L^{1/\nu}\right],
\qquad
\Delta_A=\beta_A/\nu .
\label{eq:steady_order_parameter_collapse}
\end{equation}
Here $A$ denotes the order parameter being collapsed, $p$ is the tuning probability for the corresponding protocol, and $p_c$, $\nu$, and $\beta_A$ are transition-specific. In the model-specific discussions below, we therefore specify only the diagnostic $A$ and the tuning parameter $p$ used in Eq.~\eqref{eq:steady_order_parameter_collapse}.

In addition to correlations, we examine the average halfcut bipartite entanglement entropy (EE). For a bipartition into halves $A$ and $B$, the reduced density matrix $\rho_{A}=\operatorname{Tr}_{B}[|\psi(t)\rangle \langle\psi(t)|]$ gives the halfcut Renyi EE  
\begin{equation}
S_n = \frac{1}{1-n}\log\operatorname{Tr}_{A}[\rho_{A}^n]
\label{eqn:RenyiEntropy}
\end{equation}
and the von Neumann EE  ($n=1$)
$
S_1 = -\operatorname{Tr}_{A}[\rho_{A}\log_{2}\rho_{A}].
$
At the critical points identified and in the critical phase the halfcut entanglement grows logarithmically with system size, we characterize it by the coefficient $\alpha(n=1)$, defined through
\begin{equation}
  \overline{S_n(L/2)} = \alpha(n)\ln L + C,
  \label{eq:alpha_entanglement_scaling}
\end{equation}
where $S_n(L/2)$ is the $n$th Renyi entropy at the cut at $L/2$, $C$ is a nonuniversal constant and the argument $1$ denotes the von Neumann entropy. Here $S_1$ and $\alpha(1)$ are expressed in bits. 

\subsection{NN AKLT SPT to Product-Trivial Transition}
The first route takes the NN AKLT SPT to the symmetry-preserving, featureless $|0\rangle^{\otimes L}$ product-trivial state favored by the single-site $S_i^z=0$ projector, as illustrated in Fig.~\ref{fig:AKLT_circuit_diagram}(d). The two area law endpoints of the phase diagram versus $p_{\mathrm{TT}}$ in Fig.~\ref{fig:AKLT_circuit_diagram}(g) 
are distinguished by the NN string order and bipartite-entanglement diagnostics, while the product endpoint carries no conventional local order.
\begin{figure*}[htbp]
    \includegraphics[width=7in]{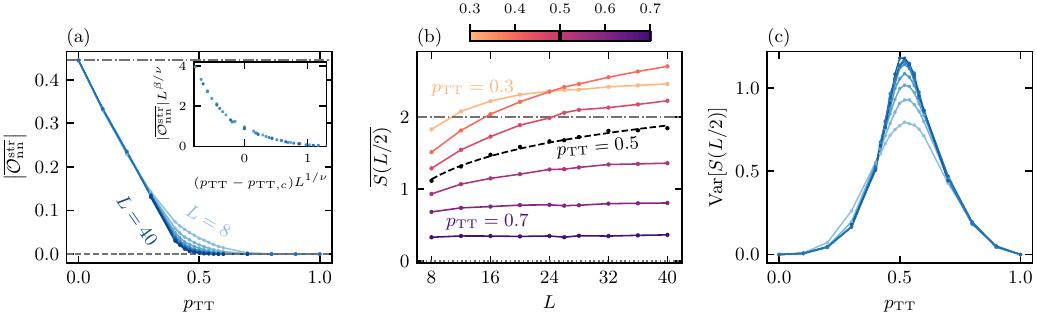}
    \caption[]{
    Data are shown for the SPT--TT model depicted in Fig.~\ref{fig:AKLT_circuit_diagram}(d), where the TT projector is applied with probability $p_{\mathrm{TT}}$ and the NN AKLT projector with probability $1-p_{\mathrm{TT}}$. (a) $\abs{\overline{\mathcal{O}_{\mathrm{nn}}^{\mathrm{str}}}}$, defined in Eq.~\eqref{eq:nn_SOP}, as a function of $p_{\mathrm{TT}}$ for available even system sizes up to $L=40$. The overline denotes averaging over circuits. The MIPT location $p_{\mathrm{TT},c}=0.50(1)$, correlation-length exponent $\nu=1.4(2)$, and order-parameter exponent $\beta=2.0(3)$ are obtained from the finite-size scaling collapse in the inset following Eq.~\eqref{eq:steady_order_parameter_collapse}. The horizontal reference lines mark the analytic limiting values of the NN string order parameter: $|\mathcal{O}_{\mathrm{nn}}^{\mathrm{str}}|=0$ in the TT product-state limit and $|\mathcal{O}_{\mathrm{nn}}^{\mathrm{str}}|=4/9$ in the NN AKLT limit~\cite{dennijsPrerougheningTransitionsCrystal1989}. (b) Scaling of $\overline{S(L/2)}$ with $L$ in the two area law phases ($p_{\mathrm{TT}}=0.30$ and $p_{\mathrm{TT}}=0.70$) and at the critical point ($p_{\mathrm{TT}}=p_{\mathrm{TT},c}=0.50(1)$). The critical data at $p_{\mathrm{TT}}=0.50$ are fit to $a\ln L+b$ by the black dashed curve. The horizontal reference line at $S(L/2)=2$ is the analytic PBC halfcut entanglement of the NN AKLT state~\cite{royBulkedgeCorrespondenceHaldane2021}. (c) Trajectory-to-trajectory variance $\operatorname{Var}[S(L/2)]$ as a function of $p_{\mathrm{TT}}$. The variance peaks at the same critical point, $p_{\mathrm{TT},c}=0.50(1)$, and provides a separate diagnostic of the transition.
    }
    \label{fig:zero_plot}
\end{figure*}
\subsubsection{String Order Transition}
The AKLT SPT phase has no symmetry-breaking long-range local order parameter: symmetry forces one-point magnetization such as $\langle S_i^z\rangle$ to vanish, while two-point spin correlations decay exponentially~\cite{affleckRigorousResultsValencebond1987,affleckValenceBondGround1988}.
We therefore use the nonlocal SOP, which was introduced to detect hidden antiferromagnetic order and is closely tied to the hidden $\mathbb{Z}_2\times \mathbb{Z}_2$ structure of the Haldane phase~\cite{dennijsPrerougheningTransitionsCrystal1989,hatsugaiNumericalStudyHidden1991,kennedyHidden221992,pollmannEntanglementSpectrumTopological2010}. The SOP is defined as
\begin{equation}
\mathcal{O}_{\mathrm{nn}}^{\mathrm{str}}(k) = \langle S_{i}^{z} \left(\prod_{j=i+1}^{i+k-1} e^{i\pi S_{j}^{z}} \right)S_{i+k}^{z}\rangle.
\label{eq:nn_SOP}
\end{equation}
We calculate $\mathcal{O}_{\mathrm{nn}}^{\mathrm{str}}(k)$ at separation $k=L/2$ and vary the system size $L$. 
In the monitored dynamics, the magnitude $|\mathcal{O}_{\mathrm{nn}}^{\mathrm{str}}(k)|$ increases from 0 as the string order develops. We use $\mathcal{O}_{\mathrm{nn}}^{\mathrm{str}}$ as an order parameter for the topological MIPT, evaluated after evolving to the fixed final time $t=4L^{2}$.

Competition between the NN AKLT and TT projectors connects two gapped, symmetry-preserving phases: the Haldane SPT ground state of Eq.~\ref{eq:AKLT_ground_state} and the topologically trivial large-$D$ product phase favored by Eq.~\ref{eq:anisotropy}.  
Below the critical value $p_{\mathrm{TT},c}$, the steady state is area law entangled and the realization-averaged string order remains finite, approaching the NN AKLT value $\overline{\mathcal{O}_{\mathrm{nn}}^{\mathrm{str}}}=-4/9$ only as $p_{\mathrm{TT}}\to0$. Above $p_{\mathrm{TT},c}$, the TT projector produces a disentangled state with $\mathcal{O}_{\mathrm{nn}}^{\mathrm{str}}\to0$ as $L\to\infty$. Near the transition, the string order parameter vanishes as
\begin{equation}
  \big|\overline{\mathcal{O}_{\mathrm{nn}}^{\mathrm{str}}}\big|\sim |p_{\mathrm{TT}}-p_{\mathrm{TT},c}|^{\beta}, 
\end{equation} 
where $p_{\mathrm{TT},c}$ is the critical rate of topologically trivial projectors and $\beta$ is the order parameter exponent.
At the critical point $p_{\mathrm{TT},c}=0.50(1)$, the SOP decays algebraically with system size,
\begin{equation}
\big|\overline{\mathcal{O}_{\mathrm{nn}}^{\mathrm{str}}}\big| \sim L^{-\beta/\nu}, 
    \end{equation}
    consistent with scale invariance at the transition.

For each $p_{\mathrm{TT}}$ and $L$, we evaluate the string correlator at $k=L/2$ and define $\overline{\mathcal{O}_{\mathrm{nn}}^{\mathrm{str}}}(L,p_{\mathrm{TT}})\equiv\overline{\mathcal{O}_{\mathrm{nn}}^{\mathrm{str}}(k=L/2)}$. Figure~\ref{fig:zero_plot} shows that hidden order persists up to $p_{\mathrm{TT},c}=0.50(1)$. Near a continuous transition, the correlation length scales as $\xi\propto|p-p_c|^{-\nu}$ and is cut off by $L$ in a finite chain.
For the SPT--TT transition, Eq.~\eqref{eq:steady_order_parameter_collapse} is evaluated with $A=\abs{\mathcal{O}_{\mathrm{nn}}^{\mathrm{str}}}$ and $p=p_{\mathrm{TT}}$.
The collapse in Fig.~\ref{fig:zero_plot} yields $p_{\mathrm{TT},c}=0.50(1)$, $\nu=1.4(2)$, and $\beta=2.0(3)$.

\begin{figure*}[htbp]
    \includegraphics[width=7in]{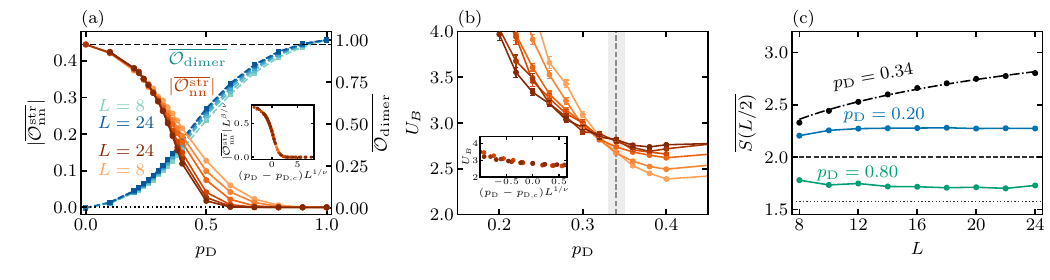}
    \caption[]{
    Data are shown for the SPT--dimer model depicted in Fig.~\ref{fig:AKLT_circuit_diagram}(e), where the dimer projector is applied on odd bonds with probability $p_{\mathrm{D}}$ and the NN AKLT projector with probability $1-p_{\mathrm{D}}$. (a) $\abs{\overline{\mathcal{O}_{\mathrm{nn}}^{\mathrm{str}}}}$, defined in Eq.~\eqref{eq:nn_SOP}, and $\overline{\mathcal{O}_{\mathrm{dimer}}}$, defined in Eq.~\eqref{eq:dimer_op}, as functions of $p_{\mathrm{D}}$ for $L\in\{8,10,12,16,20,24\}$. The overline denotes averaging over circuits. The horizontal reference lines mark the analytic limiting values of the two order parameters. The inset in panel (a) shows a finite-size scaling collapse of $\abs{\overline{\mathcal{O}_{\mathrm{nn}}^{\mathrm{str}}}}$ following Eq.~\eqref{eq:steady_order_parameter_collapse}, yielding $p_{\mathrm{D},c}=0.34(1)$, $\nu=1.3(2)$, and $\beta=0.22(3)$. (b) Centered Binder cumulant $U_B$, defined in Eq.~\eqref{eq:dimer_binder}, for the dimer order parameter as a function of $p_{\mathrm{D}}$. Error bars are delete-one-trajectory jackknife standard errors, with all retained late-time samples from a trajectory treated as one block. The inset shows the collapse of $U_B$ for $L\in\{18,20,24\}$, using $p_{\mathrm{D},c}=0.34$ and $\nu=1.3$ from panel (a). The dashed line and gray band show the central value and uncertainty from panel (a), respectively.
    (c) Scaling of $\overline{S(L/2)}$ with $L$ in the two area law phases ($p_{\mathrm{D}}=0.20$ and $p_{\mathrm{D}}=0.80$) and at the critical point ($p_{\mathrm{D}}=p_{\mathrm{D},c}=0.34(1)$). The critical data at $p_{\mathrm{D}}=0.34$ are fit to $a\ln L+b$ by the black dashed curve. The horizontal reference lines at $S(L/2)=2$ and $S(L/2)=\log_2 3$ correspond to the analytic PBC halfcut entropy of the NN AKLT state~\cite{royBulkedgeCorrespondenceHaldane2021} and to a cut through one spin-1 singlet dimer, respectively.
    }
    \label{fig:dimer_plot}
\end{figure*}

\subsubsection{Entanglement Scaling}

Both sides of the SPT--TT transition are governed by area law wavefunctions. At the critical point, the halfcut entanglement instead grows logarithmically with $L$, and the dynamical collapse in Appendix~\ref{sec:dynamical_exponent} gives $z=1.16(5)$.
We examine the halfcut bipartite entanglement entropy (EE) and its trajectory-to-trajectory variance as functions of $p_{\mathrm{TT}}$. 
$\overline{S(L/2)}$ remains approximately independent of system size on both sides of the transition, consistent with area law entanglement in the SPT and TT phases, as shown in Fig.~\ref{fig:zero_plot}(b). At criticality, for $p_{\mathrm{TT}}=p_{\mathrm{TT},c}=0.50(1)$, the entanglement instead scales logarithmically with $L$, as shown by the dashed black line in Fig.~\ref{fig:zero_plot}(b). In the NN AKLT limit, the PBC halfcut bipartition cuts two virtual AKLT bonds, so the halfcut entropy approaches $\overline{S(L/2)}=2$, as shown in Fig.~\ref{fig:zero_plot}(b).
At criticality, the dynamical collapse gives $z=1.16(5)$ for this transition, as reported in Table~\ref{tab:single_layer_cluster} and Appendix~\ref{sec:dynamical_exponent}. Thus, TT projectors destroy the hidden string order through a continuous measurement-induced phase transition, without generating conventional symmetry-breaking order. 

We also examine the trajectory-to-trajectory variance $\operatorname{Var}[S(L/2)]$ as an independent indicator of the transition, following the operator-scaling analysis of measurement-induced criticality~\cite{zabaloOperatorScalingDimensions2022}. We compute the mean and variance from the steady-state values $S(t=4L^{2})$ across independent trajectories, after the initial entanglement growth has saturated. Figure~\ref{fig:zero_plot}(c) shows $\operatorname{Var}[S(L/2)]$, with darker curves denoting larger system sizes. The variance is expected to be maximal near the critical measurement rate $p_{\mathrm{TT},c}$, and the data show a clear maximum consistent with the SOP estimate $p_{\mathrm{TT},c}=0.50(1)$. The logarithmic fit at the SPT--TT critical point in Fig.~\ref{fig:zero_plot}(b) gives $\alpha(1)=0.48(1)$, as reported in Table~\ref{tab:single_layer_cluster}. This value is close to the two-dimensional percolation reference value $\sqrt{3}/\pi\simeq0.55$, indicating relatively weak logarithmic entanglement growth at this transition.

\subsection{NN AKLT SPT to Explicitly Dimerized VBS Transition}
In the second model, the NN AKLT projector competes with the odd-bond dimer projector shown in Fig.~\ref{fig:AKLT_circuit_diagram}(e). The projector pattern explicitly breaks translation symmetry. String order, dimer order, and bipartite entanglement distinguish the NN AKLT regime at $p_{\mathrm{D}}<p_{\mathrm{D},c}=0.34(1)$ from the explicitly dimerized regime at $p_{\mathrm{D}}>p_{\mathrm{D},c}$.

\subsubsection{String and Dimer Order Transition}

We evolve the circuit to a steady state at $t=5L^{2}$. Compared with the SPT--TT model, the SPT--dimer VBS transition uses this longer fixed final time at criticality to obtain stable finite-size estimates of the order parameter and fitted critical parameters.

We distinguish the two phases using $\mathcal{O}_{\mathrm{nn}}^{\mathrm{str}}$, defined in Eq.~\ref{eq:nn_SOP}, and the staggered dimer order parameter
\begin{equation}
\mathcal{O}_{\mathrm{dimer}} = \frac{1}{L}\sum_{i=1}^{L} (-1)^{i} \langle \mathbf{S}_{i} \cdot \mathbf{S}_{i+1} \rangle.
\label{eq:dimer_op}
\end{equation}
As a separate diagnostic of the transition, we compute the centered Binder ratio $U_B$ of the dimer order parameter,
\begin{equation}
U_B =
\frac{\overline{(\delta \mathcal{O}_{\mathrm{dimer}})^{4}}}
{\overline{(\delta \mathcal{O}_{\mathrm{dimer}})^{2}}^{\,2}},
\label{eq:dimer_binder}
\end{equation}
where $\delta \mathcal{O}_{\mathrm{dimer}}=\mathcal{O}_{\mathrm{dimer}}-\overline{\mathcal{O}_{\mathrm{dimer}}}$ and the overline denotes averaging over circuit realizations. Curves of this dimensionless ratio for different system sizes should cross near the critical point, up to finite-size drift.
For $p_{\mathrm{D}}<p_{\mathrm{D},c}$, the NN AKLT projector favors the SPT phase with long-range string order. At $p_{\mathrm{D}}=0$, $\mathcal{O}_{\mathrm{nn}}^{\mathrm{str}}\to-4/9$ in the thermodynamic limit and $\mathcal{O}_{\mathrm{dimer}}=0$. For $p_{\mathrm{D}}>p_{\mathrm{D},c}$, the system enters the explicitly dimerized VBS selected by the measurement protocol. At $p_{\mathrm{D}}=1$, the state is a product of nearest-neighbor singlets on alternating bonds, as shown in Fig.~\ref{fig:AKLT_circuit_diagram}(e), with $\mathcal{O}_{\mathrm{dimer}}=1$ and $\mathcal{O}_{\mathrm{nn}}^{\mathrm{str}}=0$. Appendix~\ref{sec:SOP_analytic} derives the limiting string order values.

As shown in Fig.~\ref{fig:dimer_plot}(a), $\abs{\overline{\mathcal{O}_{\mathrm{nn}}^{\mathrm{str}}}}$ decreases from its NN AKLT limiting value $4/9$ toward zero as the probability $p_{\mathrm{D}}$ of applying the competing dimer projector increases, with the drop becoming sharper for larger $L$. The finite-size scaling collapse shown in the inset of Fig.~\ref{fig:dimer_plot}(a) locates the phase transition at $p_{\mathrm{D},c}=0.34(1)$, where long-range NN string order vanishes in the thermodynamic limit.

At the two pure-projector endpoints, $\abs{\overline{\mathcal{O}_{\mathrm{nn}}^{\mathrm{str}}}}\to4/9$ in the NN AKLT state and $\overline{\mathcal{O}_{\mathrm{nn}}^{\mathrm{str}}}=0$ in the dimer state.
For the SPT--dimer transition, Eq.~\eqref{eq:steady_order_parameter_collapse} is evaluated with $A=\abs{\mathcal{O}_{\mathrm{nn}}^{\mathrm{str}}}$ and $p=p_{\mathrm{D}}$.
The inset of Fig.~\ref{fig:dimer_plot}(a) shows a scaling collapse of this form yielding $p_{\mathrm{D},c}=0.34(1)$, $\nu=1.3(2)$, and $\beta=0.22(3)$.
The Binder cumulant in Fig.~\ref{fig:dimer_plot}(b) provides an independent, dimensionless check on this transition point. Since $U_B$ is constructed from the centered distribution of the dimer order parameter, its finite-size curves are expected to cross near a continuous transition, up to finite-size drift and corrections to scaling. The pairwise crossings of the various system-size curves give the estimate $p_{\mathrm{D},c}^{(U_B)}=0.32(2)$, where the uncertainty reflects their substantial finite-size drift. After omitting the smallest sizes, the crossings cluster near $p_{\mathrm{D}}=0.33$, consistent with the string order collapse.
We therefore use the Binder analysis as a complementary diagnostic of the SPT--dimer transition, while quoting the final value $p_{\mathrm{D},c}=0.34(1)$ from the finite-size collapse of $\abs{\overline{\mathcal{O}_{\mathrm{nn}}^{\mathrm{str}}}}$.

\subsubsection{Entanglement Transition}
Figure~\ref{fig:dimer_plot}(c) shows the scaling of $\overline{S(L/2)}$ across the SPT--dimer VBS transition. Both endpoints, the NN AKLT phase at $p_{\mathrm{D}}=0$ and the dimer VBS phase at $p_{\mathrm{D}}=1$, obey an area law.
\begin{figure*}[htbp]
    \includegraphics[width=7in]{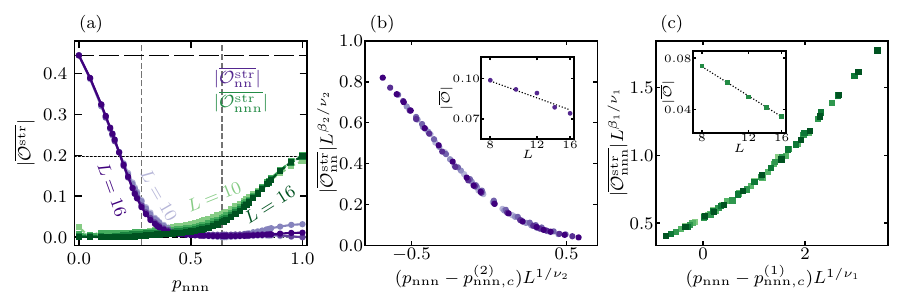}
    \caption[]{Data are shown for the NN--NNN AKLT model as depicted in Fig.~\ref{fig:AKLT_circuit_diagram}(f), where the NNN AKLT projector is applied with probability $p_{\mathrm{nnn}}$ and the NN AKLT projector with probability $1-p_{\mathrm{nnn}}$. Order-parameter diagnostics for the NN--NNN AKLT statevector transition are shown for $L\in\{8,10,12,14,16\}$.
    (a) Global view of $|\overline{\mathcal{O}_{\mathrm{nn}}^{\mathrm{str}}}|$ and $|\overline{\mathcal{O}_{\mathrm{nnn}}^{\mathrm{str}}}|$ as functions of $p_{\mathrm{nnn}}$. The NN AKLT Haldane SPT and topologically trivial NNN AKLT regimes are separated by an intermediate interval in which both measured string order parameters vanish. The entanglement scaling in Fig.~\ref{fig:nn_nnn_collapse_stack} supports the identification of this interval as an extended critical regime. Horizontal reference lines mark the analytic limiting values $0$, $4/9$, and $(4/9)^2$ from Appendix~\ref{sec:SOP_analytic}. The vertical dashed lines mark the two phase boundaries.
    (b) Finite-size collapse of $|\overline{\mathcal{O}_{\mathrm{nn}}^{\mathrm{str}}}|$ at the NN Haldane SPT boundary, using $p_{\mathrm{nnn},c}^{(2)}=0.28(1)$ and $(\nu_2,\beta_2)=(1.8(3),0.7(4))$. The inset shows the critical slice versus $L$ on log--log axes at $p_{\mathrm{nnn}}=p_{\mathrm{nnn},c}^{(2)}$.
    (c) Finite-size collapse of $|\overline{\mathcal{O}_{\mathrm{nnn}}^{\mathrm{str}}}|$ at the NNN AKLT boundary, using $p_{\mathrm{nnn},c}^{(1)}=0.64(1)$ and $(\nu_1,\beta_1)=(0.96(5),0.9(1))$. The inset shows the critical slice versus $L$ on log--log axes at $p_{\mathrm{nnn}}=p_{\mathrm{nnn},c}^{(1)}$. Dotted lines in the insets are fixed-exponent guides proportional to $L^{-\beta/\nu}$, using the $\beta$ and $\nu$ values from the corresponding collapse.
    }
    \label{fig:nn_nnn_plot}
\end{figure*}
\begin{figure*}[htbp]
    \centering
    \includegraphics[width=7in]{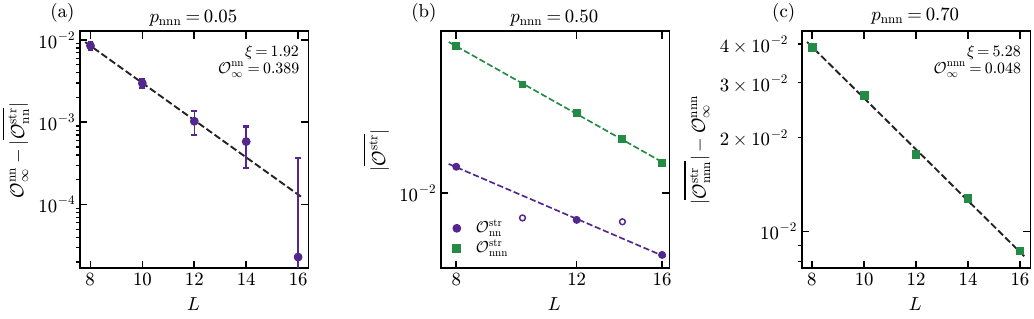}
    \caption[]{Finite-size decay of the string order parameters across the NN--NNN AKLT phase diagram. (a) In the NN AKLT phase at $p_{\mathrm{nnn}}=0.05$, the correction to the nonzero NN string order plateau decays exponentially, $\mathcal{O}_{\infty}^{\mathrm{nn}}-|\overline{\mathcal{O}_{\mathrm{nn}}^{\mathrm{str}}}|\propto e^{-L/\xi}$, with $\xi=1.92$ and $\mathcal{O}_{\infty}^{\mathrm{nn}}=0.389$. (b) At the representative critical-phase point $p_{\mathrm{nnn}}=0.50$, the NN and NNN string order parameters are shown with separate power-law fits $|\overline{\mathcal{O}_{\alpha}^{\mathrm{str}}}|\propto L^{-\eta_{\alpha}}$. 
    Filled and open NN markers distinguish the two $L\bmod 4$ branches, and the NN fit uses the filled-marker branch. 
    The same algebraic decay is observed at representative points throughout the intermediate critical phase, as shown in Fig.~\ref{fig:nn_nnn_critical_phase_decay} of Appendix~\ref{sec:nn_nnn_saturation}.
    (c) In the NNN AKLT phase at $p_{\mathrm{nnn}}=0.70$, the correction to the nonzero NNN string order plateau decays exponentially, $|\overline{\mathcal{O}_{\mathrm{nnn}}^{\mathrm{str}}}|-\mathcal{O}_{\infty}^{\mathrm{nnn}}\propto e^{-L/\xi}$, with $\xi=5.28$ and $\mathcal{O}_{\infty}^{\mathrm{nnn}}=0.048$.
    }
    \label{fig:nnn_string_decay_exponent}
\end{figure*}
In the periodic NN AKLT state, a halfcut bipartition crosses two virtual AKLT bonds and gives $\overline{S(L/2)}=2$. In the odd-bond dimer product state, the entropy is zero when the cut lies between dimers and $\log_2 3$ when it severs one spin-1 singlet. The halfcut entropy in the dimer phase therefore depends on the cut location.

At $p_{\mathrm{D},c}=0.34(1)$, $S(L/2)$ grows logarithmically with system size, as shown in Fig.~\ref{fig:dimer_plot}(c). The string order collapse, Binder cumulant, and entanglement scaling therefore identify the same transition between NN AKLT string order and explicitly selected dimer order.
The logarithmic coefficient is $\alpha(1)=0.37(2)$, smaller than the SPT--TT value and consistent with weaker critical entanglement growth at the SPT--dimer transition.

\subsection{NN AKLT SPT to Topologically Trivial NNN AKLT Transition}
In the third model, the NN AKLT projector competes with the NNN AKLT projector. The NNN AKLT endpoint is topologically trivial under the diagonal on-site $SO(3)$ symmetry preserved by the mixed protocol. 
The endpoint of $p_{\mathrm{nnn}}=1$ remains entangled and contains two interleaved AKLT structures, but its paired projective edge representations combine into a linear representation under the diagonal symmetry.
The two transitions and a critical phase are a property of this constrained competing-projector path rather than a protected distinction between two trivial endpoints.
Moreover, our attempt to simulate this model with MPS fails, and we find the bond dimension required to accurately represent the wavefunction remains exponentially growing with $L$, see Appendix~\ref{sec:bond_dimension}.
For $L\leq16$, the separated string order boundaries and logarithmic entanglement growth are consistent with an intervening extended critical regime. As $p_{\mathrm{nnn}}$ increases, $p_{\mathrm{nnn},c}^{(2)}$ separates the NN AKLT regime from this interval, and $p_{\mathrm{nnn},c}^{(1)}$ separates the interval from the NNN AKLT endpoint.
\subsubsection{String Order Transitions}

To track the hidden antiferromagnetic order of the topologically trivial NNN AKLT endpoint, we use the NNN SOP; this diagnostic characterizes the order inherited from its two interleaved AKLT chains but is not by itself the basis for its diagonal-symmetry SPT classification:
\begin{equation}
\mathcal{O}_{\mathrm{nnn}}^{\mathrm{str}}(k) = \langle S_{i}^{z} S_{i+1}^{z} \exp(i \pi \sum_{j=i+2}^{i+k-2}S_{j}^{z}) S_{i+k-1}^{z} S_{i+k}^{z}\rangle.
\label{eq:nnn_sop}
\end{equation}
We compute $\mathcal{O}_{\mathrm{nnn}}^{\mathrm{str}}(k)$ at separation $k=L/2$. This four-endpoint diagnostic factorizes into AKLT string operators on the even and odd sublattices and serves as the steady-state order parameter for the topologically trivial NNN AKLT endpoint. The corresponding order parameter for the NN AKLT state is $\mathcal{O}_{\mathrm{nn}}^{\mathrm{str}}(k)$. In the intermediate interval, both measured string orders are consistent with algebraically  vanishing as $L\to\infty$. 
As a result, we clearly find that the NN string order parameter and the NNN string order parameter vanish at two separate transitions. The NNN string order vanishes at $p_{\mathrm{nnn},c}^{(1)}=0.64(1)$, and the string order vanishes at $p_{\mathrm{nnn},c}^{(2)}=0.28(1)$.

Figure~\ref{fig:nnn_string_decay_exponent} compares the finite-size behavior in the two gapped phases and the intervening critical phase. In either gapped phase, the compatible string order approaches a nonzero thermodynamic value with an exponential finite-size correction. At $p_{\mathrm{nnn}}=0.50$, both string order parameters instead decay algebraically with $L$. The contrast between exponential corrections to nonzero plateaus and algebraic decay toward zero supports the intervening critical regime.
\subsubsection{Finite-Size Scaling Analysis}
\begin{figure*}[htbp]
    \centering
    \includegraphics[width=7in]{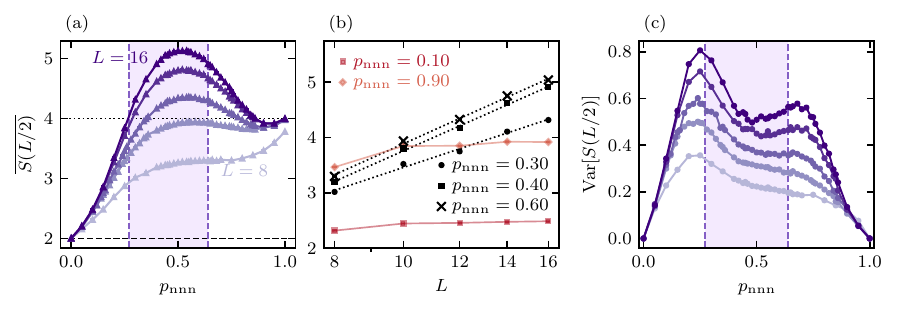}
    \caption[]{
    Data are shown for the NN--NNN AKLT model as depicted in Fig.~\ref{fig:AKLT_circuit_diagram}(f), where the NNN AKLT projector is applied with probability $p_{\mathrm{nnn}}$ and the NN AKLT projector with probability $1-p_{\mathrm{nnn}}$. Entanglement diagnostics for the NN--NNN AKLT statevector transition are shown. 
    (a) Halfcut entanglement entropy $\overline{S(L/2)}$ as a function of $p_{\mathrm{nnn}}$. The shaded region marks the interval whose finite-size signatures, at these system sizes ($L\leq16$), are consistent with an extended critical regime, as inferred from the string order collapses, and the horizontal reference lines show the exact PBC limits $S(L/2)=2$ for the NN AKLT state and $S(L/2)=4$ for the NNN AKLT state.
    (b) Scaling of $\overline{S(L/2)}$ with $L$ for $p_{\mathrm{nnn}}\in\{0.10,0.30,0.40,0.60,0.90\}$, spanning the two area law endpoints and the intervening critical regime. Over the accessible sizes, the critical-region data at $p_{\mathrm{nnn}}\in\{0.30,0.40,0.60\}$ are fit to $a\ln L+b$, shown by the black dotted curves.
    (c) Sample variance $\operatorname{Var}[S(L/2)]$ as a function of $p_{\mathrm{nnn}}$, with the same shaded interval and dashed boundary lines as in panel (a).
    }
    \label{fig:nn_nnn_collapse_stack}
\end{figure*}

For the NN--NNN problem, Eq.~\eqref{eq:steady_order_parameter_collapse} is applied separately at the two boundaries, with $p=p_{\mathrm{nnn}}$. The lower boundary uses $A=\abs{\mathcal{O}_{\mathrm{nn}}^{\mathrm{str}}}$, while the upper boundary uses $A=\abs{\mathcal{O}_{\mathrm{nnn}}^{\mathrm{str}}}$. The two NN--NNN boundaries are extracted from the finite-size collapses shown in Fig.~\ref{fig:nn_nnn_plot}. The lower boundary, where NN AKLT string order disappears, is obtained from the collapse of $\abs{\overline{\mathcal{O}_{\mathrm{nn}}^{\mathrm{str}}}}$ in Fig.~\ref{fig:nn_nnn_plot}(b), using $p_{\mathrm{nnn}}\in[0.12,0.40]$. This gives $p_{\mathrm{nnn},c}^{(2)}=0.28(1)$, $\nu_{2}=1.8(3)$, and $\beta_{2}=0.7(4)$. The upper boundary, where the system enters the topologically trivial NNN AKLT regime, is obtained from the collapse of $\abs{\overline{\mathcal{O}_{\mathrm{nnn}}^{\mathrm{str}}}}$ in Fig.~\ref{fig:nn_nnn_plot}(c), using $p_{\mathrm{nnn}}\in[0.60,0.82]$. This gives $p_{\mathrm{nnn},c}^{(1)}=0.64(1)$, $\nu_{1}=0.96(5)$, and $\beta_{1}=0.9(1)$.

At the upper NNN AKLT boundary $p_{\mathrm{nnn},c}^{(1)}$, $\mathcal{O}_{\mathrm{nnn}}^{\mathrm{str}}(k)$ decays algebraically with $L$; at the lower NN AKLT boundary $p_{\mathrm{nnn},c}^{(2)}$, $\mathcal{O}_{\mathrm{nn}}^{\mathrm{str}}(k)$ does the same. Away from criticality, $\mathcal{O}_{\mathrm{nn}}^{\mathrm{str}} \rightarrow -4/9$ in the NN AKLT phase~\cite{dennijsPrerougheningTransitionsCrystal1989}, while $\mathcal{O}_{\mathrm{nnn}}^{\mathrm{str}}$ approaches $(4/9)^{2}$ in the topologically trivial NNN AKLT regime, consistent with the analytic results in Appendix~\ref{sec:SOP_analytic}.
Within the intermediate critical phase, the finite-size data are consistent with both string order parameters vanishing as $L\to\infty$. The $p_{\mathrm{nnn}}=0.50$ data in Fig.~\ref{fig:nnn_string_decay_exponent}(b) show algebraic finite-size decay of both measured string order parameters over the accessible sizes.
For the NN--NNN protocol, we use exact statevector evolution rather than MPS evolution, so the Hilbert-space dimension grows as $3^L$ and limits the accessible sizes to $L\leq16$.
\subsubsection{Entanglement Scaling}
At both gapped endpoints, bipartite entanglement obeys area law scaling. In the NN AKLT limit ($p_{\mathrm{nnn}}=0$), the PBC halfcut crosses two virtual AKLT bonds, giving $\overline{S(L/2)}=2$. At the topologically trivial NNN AKLT endpoint ($p_{\mathrm{nnn}}=1$), the state factorizes into even- and odd-sublattice AKLT chains and the cut crosses two virtual bonds in each, giving $\overline{S(L/2)}=4$. These values match Fig.~\ref{fig:nn_nnn_collapse_stack}(a).

In the intermediate interval $p_{\mathrm{nnn}}\in[p_{\mathrm{nnn},c}^{(2)},p_{\mathrm{nnn},c}^{(1)}]$, both string order parameters vanish and $\overline{S(L/2)}$ grows logarithmically with $L$. Analogous gapless regimes with logarithmic entanglement scaling in monitored free-fermion systems have been characterized as ``metallic'' phases by analogy with Anderson transitions~\cite{poboikoMeasurementInducedPhaseTransition2024}. Together with the separated string order boundaries, these $L\leq16$ finite-size signatures are consistent with an extended critical regime along the constrained NN--NNN path~\cite{skinnerMeasurementInducedPhaseTransitions2019b,liQuantumZenoEffect2018a}, without establishing its thermodynamic persistence.

The dynamical collapses give $z$ of order unity for all four boundaries: $z=1.16(5)$ (SPT--TT), $z=1.08(3)$ (SPT--dimer), $z=1.12(8)$ (NN SPT--critical), and $z=1.08(6)$ (critical--NNN AKLT); the corresponding collapses are shown in Appendix~\ref{sec:dynamical_exponent}. As shown in Table~\ref{tab:single_layer_cluster}, $z=1$ for two-dimensional percolation, the isotropic value appropriate to conformally invariant critical percolation and monitored-circuit percolation limits~\cite{smirnovCriticalPercolationPlane2009,klockeTopologicalOrderEntanglement2022,agrawalEntanglementChargeSharpeningTransitions2022}.

The final column of Table~\ref{tab:single_layer_cluster} reports the logarithmic entanglement-scaling coefficient $\alpha(1)$, defined in Eq.~\eqref{eq:alpha_entanglement_scaling}. For the two NN--NNN AKLT boundaries, $\alpha(1)$ is taken from the closest representative point inside the intervening regime shown in Fig.~\ref{fig:nn_nnn_collapse_stack}(b). Previous work reported $\alpha(1)=1.7(2)$ for Haar-random circuits and $1.61(3)$ for Clifford circuits~\cite{zabaloCriticalPropertiesMeasurementinduced2020}, while the percolation value is $\sqrt{3}/\pi\simeq0.55$~\cite{cardyLinkingNumbersSelfAvoiding2000}. The values computed here, $0.48(1)$ and $0.37(2)$ for the NN SPT--TT and SPT--dimer transitions, respectively, are near or below the percolation value and substantially smaller than the Haar and Clifford results. By contrast, the NN--NNN values $2.2(1)$ and $2.64(7)$ are larger, indicating stronger logarithmic entanglement growth in the intervening regime. Across all parameter regimes studied, the noncritical phases remain area law entangled, while the critical transitions exhibit logarithmic, rather than volume law, entanglement scaling.

\subsubsection{Evidence of a critical phase}
We conclude this section by summarizing the evidence we have for the existence of a critical phase separating the NN and NNN AKLT wavefunctions. The transitions bounding this regime have dynamical exponents of order unity. We find $z$ close to unity throughout the critical phase (not shown) within our numerical accuracy, consistent with Lorentz invariance. We find that both the NN and NNN string order parameters vanish algebraically with $L$, at representative points throughout the intervening regime (more details are provided in Appendix \ref{sec:nn_nnn_saturation}), consistent with critical correlations. This behavior accompanies logarithmic entanglement growth. The Renyi-index dependence of the logarithmic coefficient, shown in Appendix~\ref{sec:nn_nnn_renyi_entanglement}, is inconsistent with the unitary-conformal-field-theory form and suggests the description of these critical wavefunctions will be in the form of a non-unitary conformal field theory.

\section{Discussion}
\label{sec:discussion}

Table~\ref{tab:single_layer_cluster} collects the fitted critical exponents and entanglement coefficients for all four boundaries. The critical probability follows the competing-projector convention: $p_c=p_{\mathrm{TT}}$ for SPT--TT, $p_c=p_{\mathrm{D}}$ for SPT--dimer, and $p_c=p_{\mathrm{nnn}}$ for the two NN--NNN AKLT boundaries. Parenthetical uncertainties on $p_c$, $\nu$, and $\beta$ denote half-widths of the finite-size-collapse acceptance window. For $z$, they additionally cover sensitivity to nearby temporal windows and omission of an endpoint system size; for $\alpha(1)$, they denote one standard error of the fitted log-slope.

\begin{table}[ht]
    \centering
    \caption[]{Critical probabilities and fitted exponents for the three monitored spin-1 protocols.}
    \footnotesize
    \setlength{\tabcolsep}{2.7pt}
    \resizebox{\columnwidth}{!}{\begin{tabular}{|c|c|c|c|c|c|}
     \hline
       & $p_c$ & $\nu$ &  $\beta$ &  $z$ & $\alpha(1)$ \\
      \hline\hline
      NN SPT--critical & 0.28(1) & 1.8(3) & 0.7(4) & 1.12(8) & 2.2(1) \\
     \hline
      critical--NNN AKLT & 0.64(1) & 0.96(5) & 0.9(1) & 1.08(6) & 2.64(7)\\
     \hline
      NN SPT--TT & 0.50(1) & 1.4(2) &  2.0(3) & 1.16(5) & 0.48(1)  \\
     \hline
      NN SPT--dimer VBS & 0.34(1) & 1.3(2) &  0.22(3) & 1.08(3) & 0.37(2)  \\
      \hline
      2D percolation & 0.5 & 4/3 & 5/36 & 1 & $\sqrt{3}/\pi$ \\
     \hline\hline
    \end{tabular}}
    \label{tab:single_layer_cluster}
\end{table}

The NN SPT--TT and NN SPT--dimer VBS transitions yield compatible correlation-length exponents, $\nu=1.4(2)$ and $\nu=1.3(2)$, respectively (Table~\ref{tab:single_layer_cluster}), close to the two-dimensional percolation value $\nu=4/3$ and to estimates for ordinary hybrid-circuit measurement-induced transitions. However, the fitted order-parameter exponent $\beta$ differs substantially, leaving open whether the two transitions share a full universality class. The two protocols may therefore generate comparable correlation-length growth while coupling differently to the string order parameter. 

The forced projector dynamics is idealized: it assumes that the selected projector is always applied and does not specify a physical measurement channel, correction step, or postselection probability. An experimental protocol would need postselection, conditional feedback, fusion measurements, or a weak-measurement approximation, together with a response to disallowed outcomes. Small implementation errors may shift the phase boundaries while leaving the phases visible over finite length and time scales. Uncorrected Born-rule outcomes add measurement channels that may violate the same local constraints and can destabilize or round the transitions. Symmetry-breaking noise can also suppress string order and cut off the correlation length, whereas symmetry-preserving errors may primarily shift the critical probabilities. Unlike protocols that require a particular resource state, the forced dynamics can begin from any initial state with nonzero overlap with the relevant target sector.

\section{Conclusion}
\label{sec:conclusion}

We have studied three competing-projector transitions out of the spin-1 AKLT Haldane SPT phase: to a symmetry-preserving product state, an explicitly dimerized VBS, and a topologically trivial NNN AKLT endpoint. Each transition is controlled by the relative probability of two noncommuting local projectors. The critical points show logarithmic entanglement scaling with $z\approx1$, while the noncritical phases remain area law entangled. For the NN--NNN protocol, the $L\leq16$ data are also consistent with an extended critical regime; larger systems are needed to determine whether it persists.

Figure~\ref{fig:bond_dimen} in Appendix~\ref{sec:bond_dimension} shows different costs over the accessible sizes. At the SPT--TT and SPT--dimer transitions, the fitted bond dimension grows algebraically as $\overline{\chi}\propto L^{1.35}$ and $L^{2.05}$, respectively. In the NN--NNN critical regime, it grows more rapidly, as $\overline{\chi}\propto e^{0.574L}$. The NN--NNN protocol is therefore the most demanding of the three within the simulated range, motivating exact statevector calculations at small $L$. These fits characterize the chosen MPS algorithm, truncation threshold, and observation window.

Implementing the competing-projector protocol on quantum hardware first requires a spin-1 encoding. Trapped-ion qudit processors already support the single-site spin-1 gates, projective measurements, and string order diagnostics used to prepare and characterize an AKLT ground state~\cite{edmundsConstructingSpin1Haldane2025}. Extending this platform to the present protocol requires two noncommuting projectors applied at a tunable relative rate. On qubit platforms, each spin-1 can instead be encoded in the symmetric triplet subspace of two qubits by projecting out the singlet, as demonstrated for AKLT-state preparation on a superconducting processor~\cite{chenHighfidelityRealizationAKLT2023}. This encoding doubles the physical qubit count per site but can use the larger system sizes available on current qubit hardware.

A second requirement is to remove the postselection assumed by the idealized dynamics. Each projector could be implemented as a measurement followed by conditional feedback, with an unwanted outcome corrected rather than discarded. Large adaptive circuits combining repeated measurement, classical feedback, and qubit reset have already been demonstrated~\cite{pokharelOrderChaosAdaptive2025a}. The remaining tasks are to construct correction unitaries for the projectors used here and determine how imperfect feedback shifts the phase boundaries.

Critical wavefunctions remain difficult to prepare experimentally despite their applications to quantum simulation and quantum-enhanced sensing~\cite{guoFasterStatePreparation2021,frerotQuantumCriticalMetrology2018,alamLearningDynamicQuantum2024}. 
The intervening critical regime also permits the preparation of finite-size critical states. The characteristic relaxation time scales as $t^{*}\sim L^{z}$, with $z$ close to unity and $t$ measured in local projector applications. The corresponding number of sweeps is $r^{*}=t^{*}/L\sim L^{z-1}$ and varies only weakly with size over the accessible range. 

\begin{acknowledgments}
This work was partially supported by Army Research Office Grant No.~W911NF-23-1-0144 (K.A. and J.H.P.) and U.S. Office of Naval Research Grant No.~N00014-23-1-2357 (H.P. and J.H.P.). We are grateful to the OSG Consortium~\cite{Sfiligoi2009Pilot,PordesOpen2007,OSG,OSPool} for providing computational resources for the simulations. This work was performed in part at the Aspen Center for Physics, supported by National Science Foundation Grant No.~PHY-2210452 (J.H.P.), and at the Kavli Institute for Theoretical Physics (KITP), supported by National Science Foundation Grant No.~PHY-2309135 (J.H.P.). J.H.P. acknowledges the hospitality of the International Centre for Theoretical Sciences (ICTS), Bangalore, India, through its associateship program.
\end{acknowledgments}

\bibliography{AKLT_Bibliography}
\clearpage
\appendix
\section{Dynamical Exponent $z$}
\label{sec:dynamical_exponent}
\begin{figure*}[htbp]
    \centering
    \includegraphics[width=7in]{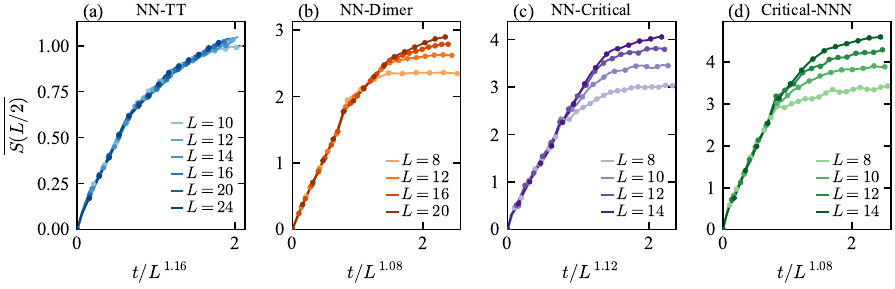}
    \caption{Dynamical scaling of the halfcut entanglement entropy. Here $t$ counts local projector applications, and each panel is plotted against the rescaled time $t/L^z$. The curves begin at $t=0$ and are displayed through $t=3L$; data up to $t=L$ are used for the collapse. At later times, the curves for different system sizes separate. Within each panel, color intensity increases with system size.
    (a) SPT--TT model at $p_{\mathrm{TT}}=0.50$ with $z=1.16(5)$, shown for $L\in\{10,12,14,16,20,24\}$.
    (b) SPT--dimer model at $p_{\mathrm{D}}=0.34$ with $z=1.08(3)$, shown for $L\in\{8,12,16,20\}$.
    (c) NN SPT--critical boundary of the NN--NNN AKLT model, $p_{\mathrm{nnn}}=0.28$, with $z=1.12(8)$, shown for $L\in\{8,10,12,14\}$.
    (d) Critical--NNN AKLT boundary, $p_{\mathrm{nnn}}=0.64$, with $z=1.08(6)$, shown for $L\in\{8,10,12,14\}$.}
    \label{fig:zero_z_entropy_collapse}
\end{figure*}
To determine the dynamical exponent $z$, we run the circuits at criticality and collapse the time-dependent quantities $\overline{A(t,L)}$, with $A$ chosen from $S(L/2)$, $\mathcal{O}_{\mathrm{nn}}^{\mathrm{str}}$, or $\mathcal{O}_{\mathrm{nnn}}^{\mathrm{str}}$. 
The temporal collapse uses the same finite-size scaling logic as Eq.~\ref{eq:steady_order_parameter_collapse}, but with the static scaling variable $(p-p_{c})L^{1/\nu}$ replaced with the temporal scaling variable $t/L^{z}$. Thus, the collapse form is
\begin{equation}
\overline{A(t,L)} \sim L^{\kappa} f_{A}(t/L^{z}),
\end{equation}
where $\kappa=\beta/\nu$ if the observable carries the order-parameter scaling dimension. Otherwise $\kappa=0$. The temporal collapses in Fig.~\ref{fig:zero_z_entropy_collapse} give dynamical exponents of order unity.
\section{Convergence of Fidelity}
\label{sec:random_product_fidelity}
\begin{figure*}[htbp]
    \centering
    \includegraphics[width=7in]{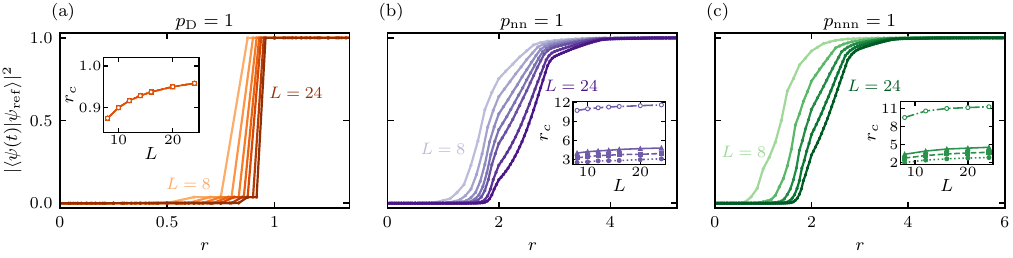}
    \caption[]{Approach to the deterministic endpoint states from random product initial states. We plot the squared fidelity $F_m(r)$ against the sweep-normalized round $r$. Panels show (a) the dimer endpoint at $p_{\mathrm{D}}=1$, (b) the NN AKLT endpoint at $p_{\mathrm{nn}}=1$, and (c) the NNN AKLT endpoint at $p_{\mathrm{nnn}}=1$. Insets show the number of rounds $r_{c}$ needed to reach the indicated fidelity thresholds: $F_m(r_{c})=0.9$ (dotted circles), $0.99$ (dashed squares), $0.999$ (solid triangles), and $1$ (dash-dotted open circles).}
    \label{fig:random_product_fidelity}
\end{figure*}

We quantify deterministic preparation at the three pure-projector endpoints described in Sec.~\ref{sec:models}: NN AKLT, NNN AKLT, and dimer. Let $t$ denote the number of local projector applications. One complete sweep contains $L$ applications, so we use the sweep-normalized round $r=t/L$. For each endpoint model $m\in\{\mathrm{nn},\mathrm{nnn},\mathrm{D}\}$, we define
\begin{equation}
F_m(r)
=
\left|
\left\langle
\psi_{\mathrm{target}}^{(m)}
\middle|
\psi_m(r)
\right\rangle
\right|^2.
\end{equation}
The three target states are
\begin{align}
\ket{\psi_{\mathrm{target}}^{(\mathrm{nn})}}
&=\ket{\mathrm{AKLT}},\\
\ket{\psi_{\mathrm{target}}^{(\mathrm{nnn})}}
&=\ket{\mathrm{AKLT}}_{\mathrm{even}}
\otimes\ket{\mathrm{AKLT}}_{\mathrm{odd}},\\
\ket{\psi_{\mathrm{target}}^{(\mathrm{D})}}
&=\ket{\psi_{\mathrm{dimer}}^{\mathrm{PBC}}}.
\end{align}
Let $\hat P^{\mathrm{nn}}$, $\hat P^{\mathrm{nnn}}$, and $\hat P^{\mathrm{D}}$ denote the NN AKLT, NNN AKLT, and dimer projectors introduced in Sec.~\ref{sec:models}. With PBC site labels understood modulo $L$, the deterministic local-slot update operators are
\begin{align}
\hat Q_{\mathrm{nn}}(a)
&=
\hat P_{a,a+1}^{\mathrm{nn}},\\
\hat Q_{\mathrm{nnn}}(a)
&=
\hat P_{a,a+2}^{\mathrm{nnn}},\\
\hat Q_{\mathrm{D}}(a)
&=
\begin{cases}
\hat P_{a,a+1}^{\mathrm{D}}, & a\ \mathrm{odd},\\
\mathbb{I}, & a\ \mathrm{even}.
\end{cases}
\end{align}
The normalized finite-time state is then
\begin{align}
\ket{\psi_{m}(r)}
&=
\mathcal{N}_{m}(r)
\hat Q_m(rL)\hat Q_m(rL-1)\cdots \hat Q_m(1)\ket{\psi_0},\\
\hat Q_m(a+L)&=\hat Q_m(a).
\end{align}
The product above contains $rL$ local updates, with $rL$ understood as an integer at the sampled rounds. The factors $\mathcal{N}_{m}(r)$ normalize the state after these updates. Neighboring local projectors generally do not commute, so the state depends on the specific projector sequence at intermediate $r$.
Figure~\ref{fig:random_product_fidelity} plots $F_{m}(r)$ for random product initial states at the three deterministic endpoints. For a fixed initial state the trajectory is deterministic, so the average is taken only over initial states.

At large $r$, each run reaches its target state with fidelity $F_{\mathrm{tot}}=1$ up to numerical precision. We also plot the mean number of rounds $r_c$ required to reach $F_{\mathrm{target}}\in\{0.9,0.99,0.999,1\}$, obtained by solving $F_m(r_c)=F_{\mathrm{target}}$. Here, one round denotes one sweep in the staircase circuit model through $L$ qutrits. The dimer state is prepared within one round for every $L$. For the NN and NNN AKLT states, $r_c$ increases moderately with $L$ but remains a few rounds over the accessible sizes. This behavior follows from the local projector constraints.

For even $L$ at $p_{\mathrm{D}}=1$, the dimer projectors act on the disjoint bonds $(1,2),(3,4),\ldots,(L-1,L)$. The identity is applied on the intervening bonds $(2,3),(4,5),\ldots,(L,1)$. Because the active projectors commute, applying each one once prepares the exact product dimer state in Eq.~\eqref{eq:dimer_projection}, provided that the initial state has nonzero overlap with the corresponding dimer state. Since each active dimer bond projector is idempotent, $\left(\hat P_{2j-1,2j}^{\mathrm{D}}\right)^2=\hat P_{2j-1,2j}^{\mathrm{D}}$, further applications do nothing. The final nontrivial update occurs on $(L-1,L)$, so $r_{c}=(L-1)/L$, as shown in Fig.~\ref{fig:random_product_fidelity}(a). To understand the origin of the $\mathcal{O}(1)$ scaling of the fidelity for the AKLT target states, we follow an analogous argument presented in~\cite{chenEfficientPreparationAKLT2024a}, where an imaginary-time evolution protocol was used to prepare the NN AKLT state. Because the AKLT projectors in the circuit act on overlapping bonds, applying one two-site AKLT filter can disturb the AKLT constraint on neighboring bonds. For example, after updating the bond $(j,j+1)$, the adjacent bonds $(j-1,j)$ and $(j+1,j+2)$ need not remain entirely outside the total-spin-$2$ sector. Equivalently, on an adjacent bond one can have
\begin{equation}
\left\langle \hat P_{i,i+1}^{S=2}\right\rangle_{\psi_m(r)} \neq 0,
\qquad i\in\{j-1,j+1\}.
\label{eq:local_spin2_disturbance}
\end{equation}
where $\hat P_{i,i+1}^{S=2}$ projects onto the total-spin-$2$ subspace of bond $(i,i+1)$. Thus a local update can temporarily violate the AKLT zero-spin-$2$ constraint on nearby bonds. However, because the AKLT state exhibits short-range correlations, this disturbance does not propagate across the entire chain. Consequently, once a round updates every local bond, the subsequent convergence is governed primarily by a local relaxation time rather than the total chain length $L$.

\section{Analytic Calculations}
\label{sec:analytic_calculations}
\subsection{Analytic Calculation of string order parameter}
\label{sec:SOP_analytic}
We calculate the finite-size dependence of the NN and NNN string order parameters in both periodic AKLT chains. The results give the horizontal reference values in the order-parameter figures and their finite-size corrections. The NN string order parameter in the NN AKLT state was previously obtained with the MPS transfer matrix~\cite{royBulkedgeCorrespondenceHaldane2021}; we use the same method for the remaining limits.

\subsubsection{Transfer-matrix setup}
The MPS of a periodic AKLT state is
\begin{equation}
|\psi\rangle_{\mathrm{AKLT}} = \sum_{i_{1}, \ldots, i_{N}} \mathrm{Tr}(A_{i_{1}} \cdots A_{i_{N}}) |i_{1} \cdots i_{N} \rangle
\label{eq:aklt_mps_periodic}
\end{equation}
where $i\in\{+1,0,-1\}$ labels the standard spin-1 basis state on each physical site, and
\begin{equation}
A_{+1}=\sqrt{\frac{2}{3}}\sigma^{+},\qquad
A_{0}=-\sqrt{\frac{1}{3}}\sigma_{z},\qquad
A_{-1}=-\sqrt{\frac{2}{3}}\sigma^{-}.
\end{equation}
$\sigma_{z}$ is the $z$ component of the Pauli spin vector, and $\sigma^{+}$ ($\sigma^{-}$) is the spin-raising (spin-lowering) operator. This representation of the AKLT state satisfies the right and left canonical conditions $\sum_{k}A_kA_k^{\dagger}=I$ and $\sum_{k}A_k^{\dagger}A_k=I$, respectively. We then define the transfer matrix
\begin{equation}
E = \sum_{k} A_k \otimes A_k^{*}.
\end{equation}
The transfer matrix is diagonalizable and can be written in terms of its eigenvectors as $E = \sum_i \lambda_{i} |e_{i}\rangle \langle e_{i}|$.
The eigenvectors are 
\begin{equation}
\label{eq:transfer_matrix_eigenvectors}
\begin{aligned}
|e_{0}\rangle &= \tfrac{1}{\sqrt{2}}\big(|00\rangle + |11\rangle\big),&
|e_{1}\rangle &= \tfrac{1}{\sqrt{2}}\big(|00\rangle - |11\rangle\big), \\
|e_{2}\rangle &= |01\rangle,&
|e_{3}\rangle &= |10\rangle,
\end{aligned}
\end{equation}
with eigenvalues
\begin{equation}
\lambda_0=1,\qquad
\lambda_1=\lambda_2=\lambda_3=-\frac{1}{3}.
\label{eq:TransferMatrix1}
\end{equation}
We denote by $Z_L$ the periodic MPS normalization,
\begin{equation}
Z_L=\Tr(E^L)=1+3\left(-\frac{1}{3}\right)^L.
\end{equation}
The subleading eigenvalue controls the finite-size corrections, with AKLT correlation length
\begin{equation}
\xi_{\mathrm{AKLT}}=\frac{1}{\ln 3}.
\end{equation}
This exact correlation length follows from the AKLT transfer
matrix~\cite{affleckRigorousResultsValencebond1987,affleckValenceBondGround1988}.

We also use the transfer-matrix identity~\cite{romanMatrixProductApproach1998}
\begin{equation}
\begin{aligned}
\mathcal{C}_{\psi}
&:=
\left\langle
\prod_{\mu=1}^{r}\mathcal{O}_{n_{\mu}}^{(\mu)}
\right\rangle_{\psi},
\\
\mathcal{C}_{\psi}
&=
\frac{1}{Z_L}\Tr\!\left[\mathcal{T}\right],
\\
\mathcal{T}
&:=
E^{n_{1}-1}\widetilde{O}^{(1)}
\prod_{\mu=2}^{r}
E^{n_{\mu}-n_{\mu-1}-1}\widetilde{O}^{(\mu)}
E^{L-n_r}.
\end{aligned}
\label{eq:transfer_matrix_identity}
\end{equation}
where the product defining $\mathcal{T}$ is ordered from left to right in increasing $\mu$. Here $\mathcal{O}_{n_{i}}^{(i)}$ is a nonidentity single-qutrit operator acting on qutrit $i$, and $1 \leq n_{1} \leq n_{2} \leq \cdots \leq n_{r} \leq L$. We also define
\begin{equation}
\tilde{O} = \sum_{k,k'} A_{k} \otimes A_{k'}^* \langle k|\mathcal{O}|k'\rangle.
\end{equation}
Here $\tilde{O}$ is the transfer-matrix insertion associated
with a local physical operator $\mathcal{O}$. We use this construction below
to represent the endpoint $S^z$ operators and the intervening string operator
in Eq.~\eqref{eq:NNN_eq1}.

\subsubsection{NNN string order in the NN AKLT state}
We start by computing the NNN string order parameter in the NN AKLT state, which is
\begin{equation}
\begin{aligned}
\label{eq:NNN_eq1}
\mathcal{O}_{\mathrm{nnn}|\mathrm{nn}}^{\mathrm{str}}(k,L) 
  &= \frac{\langle S_{k+4}^{z} S_{k+3}^{z} 
     \exp\!\left\{-\sum_{j=3}^{k+2} i \pi S_{j}^{z} \right\} 
     S_{2}^{z} S_{1}^{z} \rangle}
     {1 + 3\left(-\tfrac{1}{3}\right)^{L}} \\
  &= \frac{\mathrm{Tr}\!\left[E^{L-k-4} 
     \bigl(\tilde S^{z}\bigr)^2 
     \bigl(e^{-i\pi \tilde S^{z}}\bigr)^{k} 
     \bigl(\tilde S^{z}\bigr)^2\right]}
     {1 + 3\left(-\tfrac{1}{3}\right)^{L}} .
\end{aligned}
\end{equation}
The second line of Eq.~\eqref{eq:NNN_eq1} follows by applying the transfer-matrix identity in Eq.~\eqref{eq:transfer_matrix_identity} to the four endpoint $S^{z}$ insertions and the $k$ string operators.
We use the expansion
\begin{equation}
\label{eq:TransferMatrix2}
e^{-i\pi \tilde S^{z}} = \sum_i \lambda_i \, |\tilde e_i\rangle \langle \tilde e_i|,
\end{equation}
with the same eigenvalues as in Eq.~\eqref{eq:TransferMatrix1}. The eigenstates are related as:
\begin{equation}
\begin{array}{ll}
|\tilde{e}_0\rangle = |e_{1}\rangle & |\tilde{e}_{1}\rangle = |e_{0}\rangle \\
|\tilde{e}_{2}\rangle = |e_{2}\rangle & |\tilde{e}_{3}\rangle = |e_{3}\rangle. 
\end{array}
\end{equation}

Substituting the expansion into Eq.~\eqref{eq:NNN_eq1}, we see that 
\begin{equation}
\begin{aligned}
\label{eq:NNN_eq2}
\mathcal{O}_{\mathrm{nnn}|\mathrm{nn}}^{\mathrm{str}}(k,L) 
  &= \sum_{m,s} 
      \frac{\lambda_{m}^{\,L-k-4} \lambda_{s}^{\,k} 
      \left|\langle e_{m}| \bigl(\tilde{S}^{z}\bigr)^2 |\tilde{e}_{s}\rangle\right|^{2}}
      {1 + 3\left(-\tfrac{1}{3}\right)^{L}} .
\end{aligned}
\end{equation}
The relevant operator actions are
\begin{equation}
\begin{array}{ll}
S^{z}|e_{0}\rangle = \tfrac{2}{3}|\tilde{e}_{0}\rangle & \quad S^{z}|e_{1}\rangle = -\tfrac{2}{3}|\tilde{e}_{1}\rangle \\[6pt]
S^{z}|e_{2}\rangle = 0 & \quad (S^{z})^{2}|e_{3}\rangle = 0.
\end{array}
\end{equation}
Substituting these relations into \eqref{eq:NNN_eq1}, we find
\begin{equation}
\begin{aligned}
\mathcal{O}_{\mathrm{nnn}|\mathrm{nn}}^{\mathrm{str}}(k,L)
&=
\frac{\left(\tfrac{4}{9}\right)^2}{1+3\lambda_1^L}
\left(\lambda_1^k+\lambda_1^{L-k-4}\right) \\
&=
\left(\frac{4}{9}\right)^2(-1)^k \\
&\quad\times
\frac{
e^{-k/\xi_{\mathrm{AKLT}}}
+e^{-(L-k-4)/\xi_{\mathrm{AKLT}}}
}{
1+3e^{-L/\xi_{\mathrm{AKLT}}}
}.
\end{aligned}
\label{eq:nnn_in_nn_exact_decay}
\end{equation}
For even $L$, substituting
$\lambda_1=-1/3=-e^{-1/\xi_{\mathrm{AKLT}}}$ gives the second equality. It makes explicit the exponential decay along the two
directions around the periodic chain, while $(-1)^k$ gives the staggered sign.
When $k/L$ is held fixed as $L\to\infty$, both contributions are exponentially
suppressed, and therefore
\begin{equation}
\mathcal{O}_{\mathrm{nnn}|\mathrm{nn}}^{\mathrm{str}}(k,L) \to 0 .
\end{equation}
For the halfcut geometry used in the main figures, $k\sim L/2$, so the
residual finite-size contribution scales as
\begin{equation}
\left|\mathcal{O}_{\mathrm{nnn}|\mathrm{nn}}^{\mathrm{str}}\right|
=O(3^{-L/2}).
\end{equation}
If instead $k$ is held fixed as $L\to\infty$, the leading contribution is the
short-distance term $A_{1}(k)e^{-k/\xi_{\mathrm{AKLT}}}$.

\subsubsection{NN string order in the NNN AKLT state}
The NNN AKLT state factorizes into independent AKLT chains on the even and
odd sublattices,
\begin{equation}
|\mathrm{NNN\ AKLT}\rangle
=
|\mathrm{AKLT}\rangle_{\mathrm{even}}
\otimes
|\mathrm{AKLT}\rangle_{\mathrm{odd}} .
\label{eq:nnn_aklt_factorization}
\end{equation}
For odd endpoint separation $k$ (and hence an even number $k-1$ of
intervening sites), this factorization separates the NN string order
parameter into even- and odd-sublattice contributions:
\begin{align}
\mathcal{O}_{\mathrm{nn}|\mathrm{nnn}}^{\mathrm{str}}(k,L)
&=F_{\mathrm{even}}(k,L)F_{\mathrm{odd}}(k,L), \\
F_{\mathrm{even}}(k,L)
&=
\left\langle
S_{k+1}^{z}
\exp\!\left[-i\pi\sum_{m=1}^{(k-1)/2}S_{2m}^{z}\right]
\right\rangle_{\mathrm{even}}, \\
F_{\mathrm{odd}}(k,L)
&=
\left\langle
S_1^{z}
\exp\!\left[-i\pi\sum_{m=1}^{(k-1)/2}S_{2m+1}^{z}\right]
\right\rangle_{\mathrm{odd}}.
\end{align}
For even $k$, both endpoints lie on the same sublattice. The factor on that
sublattice is an AKLT string correlator, whereas the other factor is an
endpoint-free parity string. Its leading nonzero transfer-matrix contribution
decays as $3^{-\min(k,L-k)/2}$.
Applying Eq.~\eqref{eq:transfer_matrix_identity} separately to the two sublattices, and then using Eqs.~\eqref{eq:TransferMatrix1} and~\eqref{eq:TransferMatrix2}, shows that the odd-$k$ factors vanish, whereas the endpoint-free parity string produces the exponentially small even-$k$ branch:
\begin{equation}
\mathcal{O}_{\mathrm{nn}|\mathrm{nnn}}^{\mathrm{str}}(k,L)
=
\begin{cases}
0,
& k\ \text{odd},\\[8pt]
O\!\left(3^{-\min(k,L-k)/2}\right),
& k\ \text{even}.
\end{cases}
\end{equation}

For the halfcut separation $k=L/2$ used in the main figures, the
thermodynamic endpoint limits and parity-dependent finite-size corrections are
summarized as
\begin{equation}
\begin{array}{c|cc}
\text{diagnostic} & \text{NN AKLT} & \text{NNN AKLT} \\
\hline
\left|\mathcal{O}_{\mathrm{nn}}^{\mathrm{str}}\right|
& \frac{4}{9}+O(3^{-L})
& \substack{0\quad(k\ \mathrm{odd})\\ O(3^{-L/4})\quad(k\ \mathrm{even})} \\
\left|\mathcal{O}_{\mathrm{nnn}}^{\mathrm{str}}\right| & O(3^{-L/2}) & \left(\frac{4}{9}\right)^2+O(3^{-L/2})
\end{array}
\label{eq:SOP_endpoint_table}
\end{equation}
Thus, the NN string order in the NNN AKLT state is exactly zero at finite $L$
for odd $k$, while the even-$k$ branch vanishes exponentially as $L$ grows.
The thermodynamic limits are the reference values used in
Fig.~\ref{fig:nn_nnn_plot}. The absolute value is taken in the figures, so the
alternating signs from powers of $\lambda=-1/3$ are not visible. Finally, the
topologically trivial product state and the odd-bond dimer product state do not
possess long-range AKLT string order, giving the endpoint value
 $\mathcal{O}_{\mathrm{nn}}^{\mathrm{str}}\to 0$ used for the SPT--TT and
 SPT--dimer figures.

\subsection{Analytic Calculation of Entanglement}
\label{sec:entanglement_analytic}
{
We now derive the periodic-chain halfcut entanglement values used as
reference limits in Fig.~\ref{fig:nn_nnn_collapse_stack}(a). Throughout this subsection, entropies are measured in bits and all logarithms are base 2. The entanglement spectrum of a contiguous
block in the periodic NN AKLT state follows exactly from the MPS transfer
matrix~\cite{royBulkedgeCorrespondenceHaldane2021}. Let $L$ be the length of
the periodic AKLT chain, let $l$ be the length of the block, and define
$\gamma=-1/3$. The transfer-matrix eigenvectors $|e_\mu\rangle$ are given in
Eq.~\eqref{eq:transfer_matrix_eigenvectors}. Here, $e_\mu$ without a ket
denotes the corresponding $2\times2$ matrix obtained by reshaping
$|e_\mu\rangle=\operatorname{vec}(e_\mu)$. For an open segment of length $r$,
the physical state associated with channel $\mu$ is
\begin{equation}
|\Phi_{\mu}(r)\rangle
=
\sum_{m_1,\ldots,m_r}
\Tr\!\left[
e_{\mu}^{\dagger}A^{m_1}\cdots A^{m_r}
\right]
|m_1,\ldots,m_r\rangle .
\label{eq:aklt_segment_state}
\end{equation}
For a bipartition into a block $A$ of length $l$ and its complement $B$ of
length $L-l$, the normalized periodic AKLT state can be written in Schmidt
form as
\begin{equation}
|\Psi_{\mathrm{AKLT}}^{\mathrm{PBC}}\rangle
=
\sum_{\mu=0}^{3}
\sqrt{\lambda_{\mu}(l,L)}
|\phi_{\mu}^{A}(l)\rangle
|\phi_{\mu}^{B}(L-l)\rangle,
\end{equation}
where the segment-state overlap matrix is
\begin{equation}
G_{\mu\nu}(r)
=
\langle \Phi_{\mu}(r)|\Phi_{\nu}(r)\rangle.
\end{equation}
In the transfer-matrix eigenbasis this matrix is diagonal,
\begin{equation}
G_{\mu\nu}(r)=\delta_{\mu\nu}g_{\mu}(r),
\end{equation}
with
\begin{equation}
g_{\mu}(r)=
\begin{cases}
1+3\gamma^{r}, & \mu=0,\\
1-\gamma^{r}, & \mu=1,2,3.
\end{cases}
\end{equation}
Thus, $g_\mu(r)$ is the squared norm of the unnormalized length-$r$
segment state in channel $\mu$. A periodic bipartition contains the block and
its complement, so the unnormalized squared Schmidt coefficient is
$g_\mu(l)g_\mu(L-l)$. The total normalization is
\begin{equation}
\mathcal{N}(l,L)
=
\sum_{\mu=0}^{3}g_{\mu}(l)g_{\mu}(L-l)
=4(1+3\gamma^{L}).
\end{equation}
The normalized segment states are
\begin{equation}
|\phi_{\mu}^{A}(l)\rangle
=
\frac{|\Phi_{\mu}(l)\rangle_{A}}{\sqrt{g_{\mu}(l)}},
\qquad
|\phi_{\mu}^{B}(L-l)\rangle
=
\frac{|\Phi_{\mu}(L-l)\rangle_{B}}{\sqrt{g_{\mu}(L-l)}}.
\end{equation}
Comparing with the Schmidt decomposition gives
\begin{equation}
\lambda_{\mu}(l,L)
=
\frac{g_{\mu}(l)g_{\mu}(L-l)}{\mathcal{N}(l,L)}.
\end{equation}
Equivalently, the four nonzero Schmidt weights are
\begin{align}
\lambda_0(l,L)
&=
\frac{(1+3\gamma^l)(1+3\gamma^{L-l})}
{4(1+3\gamma^L)},\\
\lambda_a(l,L)
&=
\frac{(1-\gamma^l)(1-\gamma^{L-l})}
{4(1+3\gamma^L)},\qquad a=1,2,3 .
\end{align}
The corresponding von Neumann entropy, in the base-2 convention used in the
figures, is
\begin{equation}
\begin{aligned}
S_{\mathrm{AKLT}}^{\mathrm{PBC}}(l,L)
&=
-\lambda_0(l,L)\log_2\!\big[\lambda_0(l,L)\big] \\
&\quad
-3\lambda_1(l,L)\log_2\!\big[\lambda_1(l,L)\big].
\end{aligned}
\end{equation}
When $l$ and $L-l$ are both large, all four Schmidt weights approach $1/4$.
Because the linear correction to the entropy cancels around this uniform
spectrum,
\begin{equation}
S_{\mathrm{AKLT}}^{\mathrm{PBC}}(l,L)
=
2+O(3^{-2l})+O(3^{-2(L-l)})+O(3^{-L}).
\end{equation}
For the periodic halfcut bipartition, $l=L/2$, so the NN AKLT entropy
approaches $2$ bits with corrections of order $3^{-L}$. The NNN AKLT state
factorizes into independent even- and odd-sublattice AKLT chains. Entanglement
is additive under this tensor product, and a physical halfcut contains
approximately half of each sublattice chain. Each sublattice therefore
contributes $2$ bits in the thermodynamic limit. The endpoint entropies and
leading finite-size corrections are therefore summarized as
\begin{equation}
\begin{array}{c|cc}
\text{diagnostic} & \text{NN AKLT} & \text{NNN AKLT} \\
\hline
S(L/2) & 2+O(3^{-L}) & 4+O(3^{-L/2})
\end{array}
\label{eq:entanglement_endpoint_table}
\end{equation}
These are the two exact endpoint limits used for the horizontal reference
lines in Fig.~\ref{fig:nn_nnn_collapse_stack}.
}

\section{Finite-Size Scaling}
\label{sec:finite_size_scaling}
We use finite-size scaling collapses to estimate the critical parameters quoted in Table~\ref{tab:single_layer_cluster}. The static exponents and the dynamical exponent $z$ are obtained from the same scaling logic, but from different cuts through the scaling form. For a diagnostic $A$, the general finite-size scaling ansatz is
\begin{equation}
\overline{A(p,t,L)}=L^{-\Delta}
F_A\!\left[(p-p_c)L^{1/\nu},t/L^z\right],
\label{eq:appendix_general_collapse}
\end{equation}
where $p_c$, $\nu$, and $z$ are properties of the transition. The scaling function $F_A$ and vertical scaling dimension $\Delta$ depend on the diagnostic being collapsed. For steady-state order-parameter data, Eq.~\eqref{eq:appendix_general_collapse} reduces to the usual collapse form in which the rescaled data $L^{\beta/\nu}\overline{A}$ fall onto a single curve when plotted against $(p-p_c)L^{1/\nu}$. For temporal collapses at criticality, the same ansatz is evaluated at $p=p_c$ and the relevant horizontal variable is $t/L^z$.

To quantify the quality of a collapse, we follow the sorted-interpolation procedure used in standard data-collapse analyses~\cite{zabaloCriticalPropertiesMeasurementinduced2020}. Let $i=1,\ldots,N$ label the averaged finite-size data points included in a given collapse window. Depending on the collapse, a point may be labeled by $(p_i,L_i)$, by $(p_i,t_i,L_i)$, or by $(t_i,L_i)$ when the tuning parameter is fixed at criticality. For a trial parameter set, each point is mapped to the corresponding collapsed coordinate. For the static collapses used to estimate $p_c$, $\nu$, and $\beta$, this gives
\begin{equation}
x_i=(p_i-p_c)L_i^{1/\nu},\qquad
\widetilde{A}_i=L_i^{\Delta}\overline{A(p_i,L_i)},
\end{equation}
with $\Delta=\beta/\nu$ for the order-parameter collapses and the appropriate value of $\Delta$ set by the chosen diagnostic. We then sort the data by the collapsed coordinate $x_i$ and relabel the sorted points so that $x_1<x_2<\cdots<x_N$. For an interior point $i$, $\widetilde{A}^{\rm int}_i$ is the value of the straight line connecting the two neighboring collapsed points, $(x_{i-1},\widetilde{A}_{i-1})$ and $(x_{i+1},\widetilde{A}_{i+1})$, evaluated at $x_i$:
\begin{equation*}
\widetilde{A}^{\rm int}_i=
\frac{x_{i+1}-x_i}{x_{i+1}-x_{i-1}}\widetilde{A}_{i-1}
+\frac{x_i-x_{i-1}}{x_{i+1}-x_{i-1}}\widetilde{A}_{i+1}.
\end{equation*}
A good collapse makes the actual rescaled value $\widetilde{A}_i$ close to this neighbor-interpolated value across the data set. The collapse objective is
\begin{equation}
O(\boldsymbol{\theta})=
\frac{1}{N}\sum_i
\frac{\left(\widetilde{A}_i-\widetilde{A}^{\rm int}_i\right)^2}
{\sigma_i^2+\left(\sigma_i^{\rm int}\right)^2},
\label{eq:collapse_objective_appendix}
\end{equation}
where $\boldsymbol{\theta}$ denotes the parameters varied in the chosen ansatz, $\sigma_i$ is the rescaled statistical uncertainty of $\widetilde{A}_i$, and $\sigma_i^{\rm int}$ is the uncertainty propagated from the neighboring points used in the interpolation. The best-fit parameter set $\boldsymbol{\theta}^{\ast}$ is obtained by minimizing this objective over the chosen fitting window.

The quoted uncertainties are determined from the near-optimal region of the collapse landscape. In the two-parameter projections shown in the collapse scans, we take the allowed region to satisfy
\begin{equation}
O(\boldsymbol{\theta}) \leq 1.3\,O(\boldsymbol{\theta}^{\ast}),
\label{eq:collapse_errorbar_appendix}
\end{equation}
with all remaining parameters fixed or profiled as in the corresponding scan. For example, the uncertainty in $p_c$ is determined from the smallest and largest values of $p_c$ among all collapse parameters satisfying Eq.~\eqref{eq:collapse_errorbar_appendix}; the uncertainty in $\nu$ is determined analogously.
\begin{figure}[htbp]
    \centering
    \includegraphics[width=3.35in]{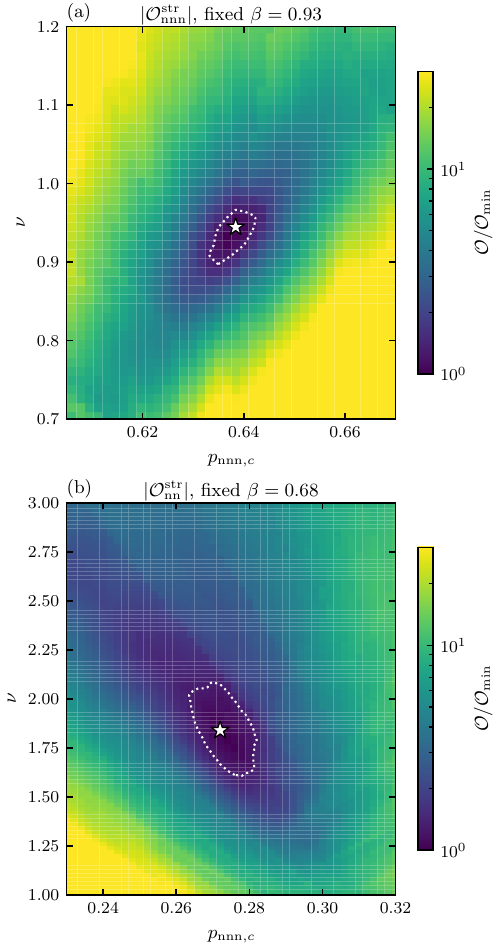}
    \caption[Goodness-of-fit landscapes for the NN--NNN AKLT statevector collapses.]{Goodness-of-fit landscapes for the NN--NNN AKLT statevector collapses, plotted as a function of $p_{\mathrm{nnn}}$. (a) Collapse objective $O(p_c,\nu)$ for $\abs{\overline{\mathcal{O}_{\mathrm{nnn}}^{\mathrm{str}}}}$ at fixed $\beta=0.9(1)$. (b) Collapse objective for $\abs{\overline{\mathcal{O}_{\mathrm{nn}}^{\mathrm{str}}}}$ at fixed $\beta=0.7(4)$. The white stars mark the grid minima, and the dotted contour marks the acceptance boundary $O=1.3O_{\min}$.}
    \label{fig:nn_nnn_additional_data}
\end{figure}
Figure~\ref{fig:nn_nnn_additional_data} shows representative objective
landscapes for the two NN--NNN AKLT string order collapses. These scans
visualize the same minimization procedure and the same
$O=1.3O_{\min}$ acceptance criterion used to quote the uncertainties in
Table~\ref{tab:single_layer_cluster}.

\section{Numerical Complexity and Bond Dimension Growth}
\label{sec:bond_dimension}
\begin{figure*}[htbp]
\centering
\includegraphics[width=7in]{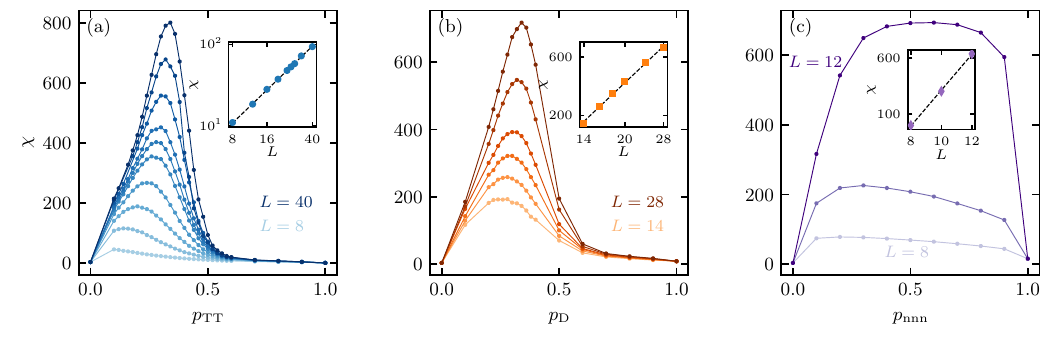}
\caption[Bond dimension growth in the three monitored spin-1 protocols.]{Bond dimension $\chi$ as a function of the measurement probability (a) $p_{\mathrm{TT}}$ in the SPT--TT model, (b) $p_{\mathrm{D}}$ in the SPT--dimer model, and (c) $p_{\mathrm{nnn}}$ in the NN--NNN AKLT model. Panel (b) uses the protocol-compatible subset $L\in\{14,16,18,20,24,28\}$ with cutoff $10^{-10}$ and the common observation window $3L\leq t\leq3L+2$; all unique compatible trajectories are retained at each point. Panel (c) uses the protocol-compatible subset $L\in\{8,10,12\}$ with cutoff $10^{-10}$, $4L\leq t\leq6L$, and three realizations per file. Insets show finite-size fits at representative values of the tuning probability: $\overline{\chi}=a_{\chi}L^{\alpha_{\chi}}$ with $\alpha_{\chi}=1.35$ at $p_{\mathrm{TT}}=0.50$ in panel (a), $\overline{\chi}=a_{\chi}L^{\alpha_{\chi}}$ with $\alpha_{\chi}=2.05$ at $p_{\mathrm{D}}=0.34$ in panel (b), and $\overline{\chi}=a_{\chi}e^{\kappa_{\chi}L}$ with $\kappa_{\chi}=0.574$ at $p_{\mathrm{nnn}}=0.50$ in panel (c).
}
\label{fig:bond_dimen}
\end{figure*}
We write the MPS wavefunction in the local basis $s_i\in\{+,0,-\}$ as
\begin{equation}
|\psi(A)\rangle = \sum_{s_{1},\ldots,s_{L}} A_{1}^{(s_{1})} A_{2}^{(s_{2})} \cdots A_{L}^{(s_{L})}|s_1s_2\cdots s_L\rangle.
\end{equation}
Here $A_{i}^{(s_{i})}$ is a $\chi_{i-1}\times\chi_i$ matrix, and the bond dimension is $\chi=\max_i\chi_i$.

We use the MPS bond dimension as a proxy for the classical simulation cost. Larger $\chi$ represents a broader Schmidt spectrum but requires more memory and computation. Figure~\ref{fig:bond_dimen} shows the trajectory-averaged bond dimension $\overline{\chi}$ as a function of the measurement probability for each protocol, using the stated MPS algorithm and truncation threshold.

The bond dimension peak should not be expected to coincide exactly with the entanglement transition. The halfcut entropy is controlled by the weighted Schmidt spectrum, whereas the MPS bond dimension at fixed truncation error is controlled by how many Schmidt values remain above the cutoff. Thus, $\overline{\chi}$ can be largest away from the transition if the dynamics produces a broader tail of Schmidt values there, even when the entropy diagnostic identifies the transition elsewhere. This distinction is especially important in monitored dynamics, where the circuit can generate broad but low-weight Schmidt spectra whose contribution to $S(L/2)$ is modest while their contribution to $\chi$ is numerically expensive.

The inset curves in Fig.~\ref{fig:bond_dimen} quantify this finite-size cost growth at representative tuning probabilities. For the SPT--TT and SPT--dimer models we fit the empirical power-law form
\begin{equation}
\overline{\chi}(L;p^{\star})=a_{\chi}L^{\alpha_{\chi}},
\label{eq:bond_dimension_power_fit}
\end{equation}
giving $\alpha_{\chi}=1.35$ at $p_{\mathrm{TT}}^{\star}=0.50$ and $\alpha_{\chi}=2.05$ at $p_{\mathrm{D}}^{\star}=0.34$. For the NN--NNN AKLT data in the protocol-compatible range $L\in\{8,10,12\}$, the curve shown in the inset is instead
\begin{equation}
\overline{\chi}(L;p_{\mathrm{nnn}}^{\star})=a_{\chi}e^{\kappa_{\chi}L},
\label{eq:bond_dimension_exp_fit}
\end{equation}
with $\kappa_{\chi}=0.574$ at $p_{\mathrm{nnn}}^{\star}=0.50$. This curve captures the rapid growth of the MPS cost near the NN--NNN mixed regime over the accessible system sizes.

\begin{table}[htbp]
\centering
\caption[]{Scaling of the bond dimension in the three monitored spin-1 protocols.}
\small
\setlength{\tabcolsep}{2.5pt}
\renewcommand{\arraystretch}{1.18}
\begin{tabular}{p{0.30\columnwidth}p{0.23\columnwidth}p{0.35\columnwidth}}
\hline\hline
Model & $p$ & $\overline{\chi}$ growth \\
\hline\hline
SPT--TT & $p_{\mathrm{TT},c}=0.50(1)$ &
$\overline{\chi}\sim L^{1.35}$ \\
& $p_{\mathrm{TT}}=0$ & $\overline{\chi}=4$ \\
& $p_{\mathrm{TT}}=1$ & $\overline{\chi}=1$ \\
\hline
SPT--dimer & $p_{\mathrm{D},c}=0.34(1)$ &
$\overline{\chi}\sim L^{2.05}$ \\
& $p_{\mathrm{D}}=0$ & $\overline{\chi}=4$ \\
& $p_{\mathrm{D}}=1$ & $\overline{\chi}=9$ \\
\hline
NN--NNN AKLT & $p_{\mathrm{nnn}}=0.50$ &
$\overline{\chi}\sim e^{0.574L}$ \\
& $p_{\mathrm{nnn}}=0$ & $\overline{\chi}=4$ \\
& $p_{\mathrm{nnn}}=1$ & $\overline{\chi}=16$ \\
\hline\hline
\end{tabular}
\label{tab:bond_dimension_growth_summary}
\end{table}

Table~\ref{tab:bond_dimension_growth_summary} summarizes the fitted growth laws and endpoint values. In the dimer model, $\overline{\chi}=4$ at $p_{\mathrm{D}}=0$ and $\overline{\chi}=9$ at $p_{\mathrm{D}}=1$. At the dimer endpoint, the maximal MPS bond splits two spin-1 singlets; each has Schmidt rank $3$, so $\chi_{\max}=3^2=9$. At intermediate probabilities, the growth is weaker than at $p_{\mathrm{D},c}$: $\overline{\chi}\sim L^{1.14}$ at $p_{\mathrm{D}}=0.20$ and $\overline{\chi}\sim L^{0.50}$ at $p_{\mathrm{D}}=0.80$. In the TT phase, $p_{\mathrm{TT}}=0.70$ gives $\overline{\chi}\sim L^{0.35}$; the NN AKLT side is also consistent with power-law growth over the available sizes.

At the NNN AKLT endpoint, $\overline{\chi}=16=2^4$ because the state factorizes into independent AKLT chains on the even and odd sublattices. A contiguous PBC partition cuts each subchain twice, crossing four virtual spin-1/2 singlets with Schmidt rank $2$. More generally, $\chi$ can grow algebraically at fixed cutoff even in an area law phase because it is sensitive to the low-weight tail of the Schmidt spectrum. The largest costs occur near the transitions or within the critical regime.

\section{Additional Results on Entanglement and Order Parameter}
\label{sec:nn_nnn_saturation}

\begin{figure*}[htbp]
    \centering
    \includegraphics[width=7in]{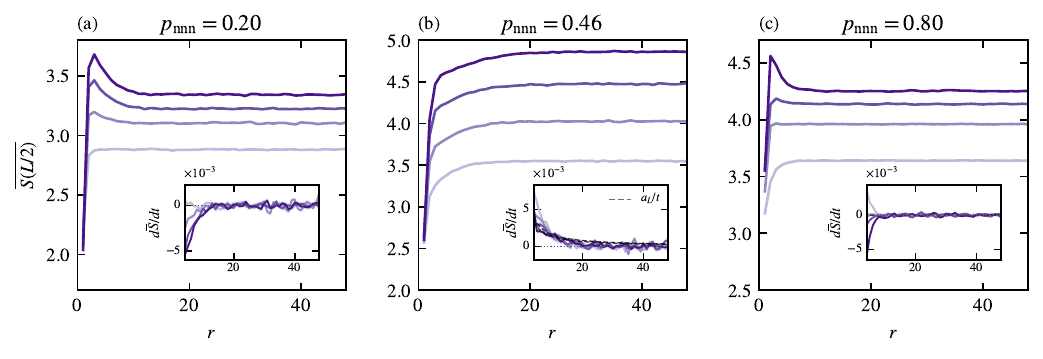}
    \caption[]{Time dependence of the halfcut entanglement
    entropy in the various phases: (a) the NN AKLT area law phase at
    $p_{\mathrm{nnn}}=0.20$, (b) the critical phase at
    $p_{\mathrm{nnn}}=0.46$, and (c) the NNN AKLT area law phase at
    $p_{\mathrm{nnn}}=0.80$. Here $t$ counts single local projector
    applications, and each panel is plotted against the rescaled time
    $r=t/L$. Insets show the corresponding time derivatives
    $d\overline{S(L/2)}/dt$; dotted horizontal lines mark zero.
    In panel (b), black dashed curves show the $a_L/t$ derivative implied by the
    logarithmic fit $\overline{S(L/2,t)}=a_L\ln t+b_L$.}
    \label{fig:nnn_saturation_entropy_time}
\end{figure*}

\begin{figure}[htbp]
    \centering
    \includegraphics[width=3.4in]{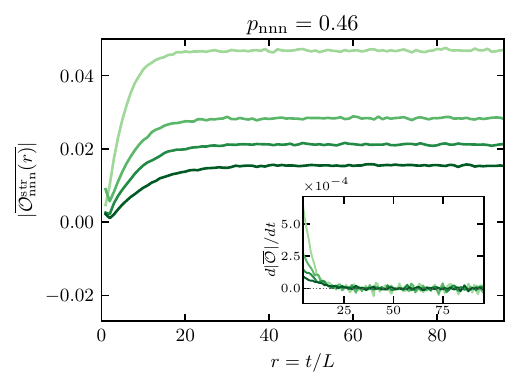}
    \caption[]{Time dependence of the trajectory-averaged NNN
    string order parameter at $p_{\mathrm{nnn}}=0.46$, a representative point
    in the critical phase. Each curve is shown at integer-round snapshots through the common final
    round $r_f=96$. The inset shows the corresponding time derivative
    $d|\overline{\mathcal{O}_{\mathrm{nnn}}^{\mathrm{str}}}|/dt$; the dotted
    horizontal line marks zero. Colors distinguish system size.}
    \label{fig:nnn_saturation_nnn_order_time}
\end{figure}

In this section, we examine the relaxation time in the critical (gapless) and
area law phases of the NN--NNN AKLT circuit model. In the area law regime, the
entropy $S(L/2)$ approaches a plateau whereas in the gapless regime, the
dynamics is compatible with logarithmic growth.
The estimated phase boundaries are $p_{\mathrm{nnn},c}^{(2)}\simeq0.28$ and
$p_{\mathrm{nnn},c}^{(1)}\simeq0.64$.

Each timestep $t$ is one local projector slot. We use the sweep-normalized
round $r=t/L$. Self-averaging over the cut locations in space is performed for
the entanglement entropy, and self-averaging over the endpoint locations is
performed for the string order parameters.

For any observable
$A\in\{S(L/2),\mathcal{O}_{\mathrm{nn}}^{\mathrm{str}},
\mathcal{O}_{\mathrm{nnn}}^{\mathrm{str}}\}$, we diagnose saturation using
the slope of the trajectory mean,
\begin{equation}
v_A(r)=\frac{d\overline{A}}{dr}
=L\frac{d\overline{A}}{dt}.
\label{eq:nnn_saturation_velocity}
\end{equation}
We estimate a saturation time from the tail of each curve, defined as the
earliest time after which the derivative persistently falls below threshold
$\epsilon$. This rules out isolated zero crossings.

\begin{figure}[htbp]
    \centering
    \includegraphics[width=3.35in]{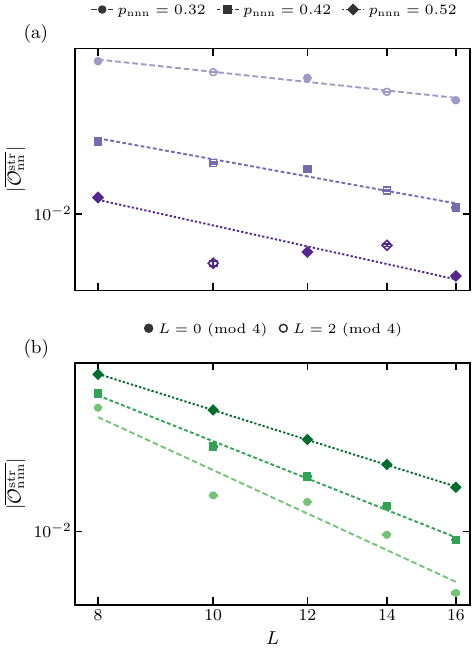}
    \caption{Finite-size algebraic decay of the string order parameters in the critical phase. Data is shown for probabilities representative of the critical phase $p_{\mathrm{nnn}}\in\{0.32,0.42,0.52\}$. (a) In the critical phase, $\abs{\overline{\mathcal{O}_{\mathrm{nn}}^{\mathrm{str}}}}\propto L^{-\eta_{\mathrm{nn}}(p_{\mathrm{nnn}})}$. The dashed line denotes the power-law fit. Filled and open NN markers distinguish the two $L\bmod 4$ branches, and the NN fits use a filled-marker branch. (b) In the critical phase, $\abs{\overline{\mathcal{O}_{\mathrm{nnn}}^{\mathrm{str}}}}\propto L^{-\eta_{\mathrm{nnn}}(p_{\mathrm{nnn}})}$.}
    \label{fig:nn_nnn_critical_phase_decay}
\end{figure}

In the area law phase, the halfcut entropy time derivative
$d\overline{S(L/2)}/dt$ shown in
the insets of Fig.~\ref{fig:nnn_saturation_entropy_time} is zero after roughly
$8$--$12$ rounds. Inside the gapless regime, the entropy continues to grow
slowly over a substantially longer window. Over the accessible intermediate
times this drift is approximately logarithmic, so that
$d\overline{S}/dt\sim1/t$, as indicated by the dashed guides in
Fig.~\ref{fig:nnn_saturation_entropy_time}(b). 

Figure~\ref{fig:nnn_saturation_nnn_order_time} shows the NNN string order
relaxation at $p_{\mathrm{nnn}}=0.46$, which lies within the critical phase.
Its finite-size value continues to drift over an
$\mathcal{O}(10)$-round transient before the derivative is zero. The saturation times are different for the entanglement
and the string order parameters. The area law observables therefore relax faster than those in the gapless regime over the accessible time window.

For $p_{\mathrm{nnn}}\in\{0.32,0.42,0.52\}$, both string order parameters decay algebraically over the accessible sizes. The NN data separate into two $L\bmod 4$ branches, and the power-law fits use the filled-marker branch. The NNN data exhibit a smoother single-branch decrease. Together, these finite-size trends are consistent with algebraic string correlations throughout the intermediate critical phase.

\section{R\'enyi-Index Dependence of Entanglement in the NN--NNN Critical Phase}
\label{sec:nn_nnn_renyi_entanglement}

\begin{figure*}[htbp]
    \centering
    \includegraphics[width=7in]{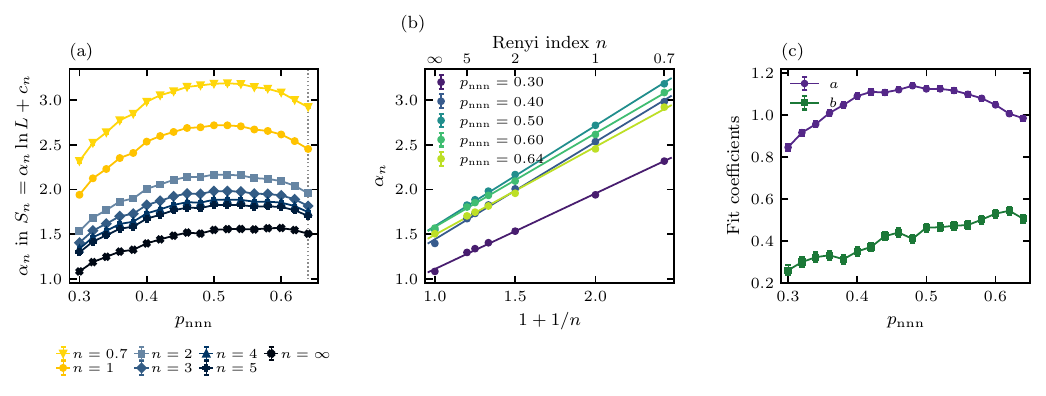}
    \caption{R\'enyi-index dependence of the logarithmic
    halfcut entanglement coefficient in the gapless critical phase of the
    NN--NNN AKLT model. (a) Coefficients $\alpha_n$ obtained from
    $S_n(L/2)=\alpha_n\ln L+c_n$ for
    $n\in\{0.7,1,2,3,4,5,\infty\}$; the vertical dotted lines mark the estimated
    phase boundaries. (b) $\alpha_n$ versus $1+1/n$ at representative values
    $p_{\mathrm{nnn}}\in\{0.30,0.40,0.50,0.60,0.64\}$. Solid curves are weighted
    fits to $\alpha_n=a(1+1/n)+b$. (c) Fitted coefficients $a$ and $b$
    across the sampled probability interval. Error bars in panels (a) and
    (b) are inherited from the finite-size fits over $L\in\{8,10,12,14\}$; those in
    panel (c) are the corresponding weighted-fit uncertainties.}
    \label{fig:nnn_renyi_alpha_summary}
\end{figure*}

We characterize the entanglement spectrum throughout the gapless critical
phase of the NN--NNN AKLT circuit using the halfcut R\'enyi entropies. For
a halfcut reduced density matrix $\rho_A$, these are
\begin{equation}
S_n=\frac{1}{1-n}\log_2\operatorname{Tr}(\rho_A^n),
\qquad
S_1=-\operatorname{Tr}(\rho_A\log_2\rho_A).
\label{eq:nnn_renyi_definition}
\end{equation}
Throughout this appendix, entropies are measured in bits, matching the
convention of Eq.~\eqref{eq:alpha_entanglement_scaling} and
Table~\ref{tab:single_layer_cluster}.

At each sampled probability, we extract the coefficient $\alpha_n$ from the
finite-size form
\begin{equation}
S_n(L/2;p_{\mathrm{nnn}})
=\alpha_n(p_{\mathrm{nnn}})\ln L+c_n(p_{\mathrm{nnn}}).
\label{eq:nnn_renyi_log_coefficient}
\end{equation}
We consider
$n\in\{0.7,1,2,3,4,5,\infty\}$. The zeroth R\'enyi entropy is omitted here because
it probes the numerically resolved Schmidt rank and consequently depends on
the threshold used to distinguish nonzero Schmidt values.

Figure~\ref{fig:nnn_renyi_alpha_summary}(a) shows strong R\'enyi-index
dependence throughout the critical phase: at every sampled
$p_{\mathrm{nnn}}$, $\alpha_n$ decreases monotonically as $n$ increases from
$0.7$ to $\infty$.

For a periodic $(1+1)$-dimensional conformal system, the standard halfcut
form is
\begin{equation}
S_n(L/2)=\frac{c}{6\ln 2}\left(1+\frac{1}{n}\right)\ln L+\mathrm{const.},
\label{eq:nnn_renyi_cft_form}
\end{equation}
which implies
\begin{equation}
\frac{\alpha_n}{\alpha_1}=\frac{1+1/n}{2}.
\label{eq:nnn_renyi_cft_ratio}
\end{equation}
The observed coefficients follow this $1+1/n$ trend qualitatively but depart
systematically from simple proportionality. Averaged over the sampled
critical probabilities, the $n=0.7$ coefficient is about $3\%$ below
Eq.~\eqref{eq:nnn_renyi_cft_ratio}; the $n=2,3,4,5,\infty$ coefficients are
about $6\%$, $9\%$, $11\%$, $12\%$, and $14\%$ above it, respectively.

We therefore fit the less restrictive empirical form
\begin{equation}
\alpha_n(p_{\mathrm{nnn}})
=a(p_{\mathrm{nnn}})\left(1+\frac{1}{n}\right)
+b(p_{\mathrm{nnn}}).
\label{eq:nnn_renyi_affine_fit}
\end{equation}
The affine form in Eq.~\eqref{eq:nnn_renyi_affine_fit} describes the data in
Fig.~\ref{fig:nnn_renyi_alpha_summary} and demonstrates the presence of non-unitary contribution, which is consistent with the critical wavefunctions being described by a non-unitary conformal field theory. Across the critical interval,
$a\simeq0.85$--$1.14$ and $b\simeq0.26$--$0.55$; the constrained form $b=0$
is strongly disfavored. Their probability dependence suggests that no single pair
$(a,b)$, or single effective central charge, describes the entire critical phase.
Thus, the leading $1+1/n$ structure requires a nonzero,
probability-dependent intercept.

\end{document}